\documentclass[12pt]{article}
\usepackage{amssymb}

\def\hybrid{\topmargin 0pt      \oddsidemargin 0pt
        \headheight 0pt \headsep 0pt
        \voffset=-0.5cm
        \textwidth 6.25in       
        \textheight 9.5in       
        \marginparwidth 0.0in
        \parskip 5pt plus 1pt   \jot = 1.5ex}
\catcode`\@=11
\def\marginnote#1{}

\newcount\hour
\newcount\minute
\newtoks\amorpm
\hour=\time\divide\hour by60 \minute=\time{\multiply\hour by60 \global\advance\minute by-\hour}
\edef\standardtime{{\ifnum\hour<12 \global\amorpm={am}%
        \else\global\amorpm={pm}\advance\hour by-12 \fi
        \ifnum\hour=0 \hour=12 \fi
        \number\hour:\ifnum\minute<10 0\fi\number\minute\the\amorpm}}
\edef\militarytime{\number\hour:\ifnum\minute<10 0\fi\number\minute}

\def\draftlabel#1{{\@bsphack\if@filesw {\let\thepage\relax
   \xdef\@gtempa{\write\@auxout{\string
      \newlabel{#1}{{\@currentlabel}{\thepage}}}}}\@gtempa
   \if@nobreak \ifvmode\nobreak\fi\fi\fi\@esphack}
        \gdef\@eqnlabel{#1}}
\def\@eqnlabel{}
\def\@vacuum{}
\def\draftmarginnote#1{\marginpar{\raggedright\scriptsize\tt#1}}
\def\draftlabel#1{{\@bsphack\if@filesw {\let\thepage\relax
   \xdef\@gtempa{\write\@auxout{\string
      \newlabel{#1}{{\@currentlabel}{\thepage}}}}}\@gtempa
   \if@nobreak \ifvmode\nobreak\fi\fi\fi\@esphack}
        \gdef\@eqnlabel{#1}}
\def\@eqnlabel{}
\def\@vacuum{}
\def\draftmarginnote#1{\marginpar{\raggedright\scriptsize\tt#1}}

\def\draft{\oddsidemargin -.5truein
        \def\@oddfoot{\sl preliminary draft \hfil
        \rm\thepage\hfil\sl\today\quad\militarytime}
        \let\@evenfoot\@oddfoot \overfullrule 3pt
        \let\label=\draftlabel
        \let\marginnote=\draftmarginnote
   \def\@eqnnum{(\theequation)\rlap{\kern\marginparsep\tt\@eqnlabel}%
\global\let\@eqnlabel\@vacuum}  }

\def\numberbysection{\@addtoreset{equation}{section}
        \def\theequation{\thesection.\arabic{equation}}}

\def\underline#1{\relax\ifmmode\@@underline#1\else
        $\@@underline{\hbox{#1}}$\relax\fi}

\def\titlepage{\@restonecolfalse\if@twocolumn\@restonecoltrue\onecolumn
     \else \newpage \fi \thispagestyle{empty}\c@page\z@
        \def\thefootnote{\fnsymbol{footnote}} }

\def\endtitlepage{\if@restonecol\twocolumn \else  \fi
        \def\thefootnote{\arabic{footnote}}
        \setcounter{footnote}{0}}  

\numberbysection \hybrid

\newcommand{\tr}{{\rm tr}}

\newtheorem{lemma}{Lemma}[section]

\newcommand{\la}{\lambda}

\newcommand{\al}{\alpha}
\newcommand{\be}{\beta}
\newcommand{\ga}{\gamma}
\newcommand{\om}{\omega}

\newtheorem{predl}{Proposition}[section]

\def\beq{\begin{equation}}
\def\eeq{\end{equation}}
\def\p{\partial}

\newtheorem{theor}{Theorem}

\newcommand{\mat}[4]{\left(\begin{array}{cc}{#1}&{#2}\\{#3}&{#4}
\end{array}\right)}

\def\res{\mathop{\hbox{Res}}\limits}

\begin{document}

\setcounter{page}{1}

\date{}
\date{}
\vspace{50mm}

\begin{flushright}
 ITEP-TH-34/13\\
\end{flushright}
\vspace{10mm}

\begin{center}
{\LARGE{Spectrum of Quantum Transfer Matrices via \\ \vskip2mm
Classical Many-Body Systems }
}\\
\vspace{8mm} {\large { A. Gorsky}\,$^{\sharp\ \natural}$\ \ \ \ { A.
Zabrodin}\,$^{\dag\ \flat\ \sharp\ \natural}$\ \ \
 { A. Zotov}\,$^{\diamond\ \sharp\ \natural}$}\\ \vspace{5mm}
 \vspace{3mm} $^\sharp$ - {\small{\em 
 ITEP, Bolshaya Cheremushkinskaya str. 25, 117218,  Moscow, Russia}}\\
 \vspace{2mm}$^\natural$ - {\small{\em MIPT, Inststitutskii per.  9, 141700, Dolgoprudny,
 Moscow region, Russia}}\\
 \vspace{2mm} $^\dag$ - {\small{\em Institute of Biochemical Physics,
Kosygina str. 4, 119991, Moscow, Russia}}\\
 \vspace{2mm} $^\flat$ - {\small{\em National Research University Higher School of
Economics,  Myasnitskaya str. 20,\\ 101000, Moscow, Russia}}\\
  \vspace{2mm} $^\diamond$ - {\small{\em Steklov Mathematical Institute, RAS, Gubkina str. 8, 119991, Moscow,
  Russia}}
\end{center}

\let\thefootnote\relax\footnote{E-mails: {\rm gorsky@itep.ru}; {\rm zabrodin@itep.ru}; {\rm zotov@mi.ras.ru}}


\begin{abstract}

In this paper we clarify the relationship between
inhomogeneous quantum spin
chains and classical integrable
many-body systems. It provides an
alternative (to the nested Bethe ansatz) method for computation of
spectra of the spin chains.
Namely, the spectrum of the quantum
transfer matrix for the inhomogeneous
${\mathfrak g}{\mathfrak l}_n$-in\-va\-ri\-ant XXX spin chain on $N$ sites
with twisted boundary conditions can be
found in terms of velocities of particles in the rational $N$-body
Ruijsenaars-Schneider model. The possible values of the velocities
are to be found from intersection points of two Lagrangian
submanifolds in the phase space of the classical model.
One of them is the Lagrangian hyperplane corresponding to
fixed coordinates of all $N$ particles and the other one is
an $N$-dimensional
Lagrangian submanifold obtained by fixing levels of $N$ classical
Hamiltonians in involution. The latter are determined
by eigenvalues of the twist matrix.
To support this picture, we give a direct proof that
the eigenvalues of the Lax matrix for the classical
Ruijsenaars-Schneider model, where velocities of particles
are substituted by eigenvalues of the spin chain Hamiltonians,
calculated through the Bethe equations, coincide with
eigenvalues of the twist matrix, with certain multiplicities.
We also prove a similar statement for the ${\mathfrak g}{\mathfrak l}_n$
Gaudin model with $N$ marked points (on the quantum side)
and the Calogero-Moser system with $N$ particles (on the classical side).
The realization of the results obtained in terms
of branes and supersymmetric gauge theories is also discussed.

\end{abstract}


\small{

\newpage

\tableofcontents

\section{Introduction}
\setcounter{equation}{0}

The notion of duality was introduced into the landscape of
integrable models long ago. In a general sense, it connects
integrable systems of different types and ranges from purely quantum
field theory models like sin-Gordon and Thirring to purely classical
integrable systems with finite number of degrees of freedom. On the
other hand, dualities play a key role in supersymmetric gauge
theories and their stringy and M-theory UV completions. In this
context, we again have plenty of correspondences between the gauge
theories like S-duality, T-duality, mirror symmetry, Seiberg duality
etc. All of them have one or another geometrical origin and many of
them can be most clearly expressed in terms of brane motion.

It is known nowadays that integrable models are closely related to
the SUSY gauge theories in different dimensions and provide an
effective tool to describe them in the low-energy sector in the
spirit of the Seiberg-Witten solution \cite{sw}. In
particular, some low-energy effective actions in N=2 gauge theories
have been obtained via mapping to the corresponding integrable
many-body systems \cite{gkmmm} (see \cite{gormir} for a review).
To some extent the
integrable systems capture the emerging hidden symmetry when the
integration over the moduli space of the non-perturbative solutions
relevant for the particular SUSY gauge theory is taken into account.
The degrees of freedom of the corresponding integrable systems as
well as the commuting flows are identified with the coordinates of the
different branes localized in  different dimensions in ten- or eleven-
dimensional geometry of the string and M-theory.

The relation between gauge theories and integrable systems can be
used in both directions. The Adams--Harnard--Hurtubise (AHH) duality
\cite{harnad2} together with the results of \cite{MTV3} allows us to
obtain an interesting interrelation \cite{zotov02} between
Heisenberg magnetic chains and their degenerate cases known as
Gaudin models \cite{Gaudin}. It was known for quite a long time that
a similar phenomenon takes place in the theory of classical
many-body systems of the Calogero-Moser (CM) \cite{Calogero} or the
Ruijsenaars-Schneider (RS) \cite{Ruijs1} types. These issues have
been discussed in \cite{ggm,braden,p-q,fock}. Recently, an improved
version of the AHH duality (the spectral duality) turned out to be a
very effective tool in analyzing the 2d/4d duality
\cite{bcgk,zotov01,zz02,bazhanov} and the AGT correspondence
\cite{agt01,ba444,ns1,ns2}. The AHH duality has been identified as
the 3d mirror symmetry \cite{gk}.

In this paper we will focus on the intriguing {\em classical-quantum (QC)
duality} between integrable
models with finite number of degrees of freedom. One of the models
is quantum and another one is classical. Let us stress that this
correspondence (based on the recent results of
\cite{Givental,MTV1,zabrodin1,zabrodin1a,zabrodin2} and older results of
\cite{zabrodin3})
has nothing to do with the quasiclassical limit.
Presumably, it exists for integrable models only.

This kind of duality suggests an
alternative way for calculating joint spectra of commuting quantum
operators (transfer matrices and Hamiltonians), without any use
of the coordinate or algebraic Bethe ansatz technique \cite{NS},
which so far was a key tool in any exact solution of quantum
integrable models with non-trivial interaction. There is also no need in such an unavoidable
intermediate step as solving Bethe equations. The spectra of {\it
quantum} Hamiltonians of an integrable system appear to be encoded
in algebraic properties of the Lax matrix for a very different and
{\it purely classical} model!

In a nutshell, the quantum spectral problem for an
 integrable spin chain on $N$ sites reformulated
 in terms of the ``QC-dual''
 $N$-body integrable 1D systems of classical mechanics
 is as follows.
 Let us fix coordinates $q_i$ of the $N$ classical particles and levels of
 the $N$ Hamiltonians $H_i$ in involution. Then possible values of particles
 velocities give spectra of the spin chain Hamiltonians.
 In other words, one may say that the eigenstates of the quantum
 Hamiltonians correspond to intersection points of two Lagrangian
 submanifolds in the $2N$-dimensional phase space of the classical
 $N$-body system.
 One of them is the $N$-dimensional hyperplane with fixed $q_i$'s and the
 other one is an $N$-dimensional Lagrangian manifold obtained by fixing
 levels of $N$ classical Hamiltonians. Since their dimensions
 are complimentary, the intersection set is a finite number of points.
 It appears that they
 contain a specific information about eigenstates of the
 quantum spin chain.

 It is natural to conjecture that it is the Yang-Yang (YY)
 function (yielding solutions
 of the Bethe equations as its critical points) that characterizes
 the structure
of the intersection set. It was argued  in \cite{nrs} that
the YY
function plays the role of the generating function for the
Lagrangian submanifold
in the classical model. However, the meaning of this statement
is still to be clarified on particular examples.

The QC duality is traced back to paper \cite{Givental}, where joint
spectra of some finite-dimensional operators were linked to the
classical Toda chain.
Later it was extended to the following cases:
\begin{itemize}
\item[a)] The Gaudin model (from the quantum side)
and the CM many-body system (from the classical side)
\cite{MTV1,zabrodin2};
\item[b)] Inhomogeneous spin chains of the $XXX$-
and $XXZ$-type with
twisted boundary conditions
(from the quantum side) and rational or trigonometric RS
many-body systems (from the classical side)
\cite{zabrodin1,zabrodin1a}
\end{itemize}

\noindent The particles coordinates in the CM or RS models were
identified with the inhomogeneities at the sites in the spin chains
while eigenvalues of the Lax matrix for the CM or RS models were
shown to coincide with eigenvalues of the twist matrix at the spin
chain side, with certain multiplicities. The QC duality has been
recently discussed in the brane framework in \cite{gk} and was
related to the duality between quiver 3d theory and 4d theory at the
interval with nontrivial boundary conditions. Moreover, it was
suggested that the  generalized duality holds when the number of
inhomogeneities in the spin chain does not coincide with the number
of particles on the classical side. However, the arguments in favour
of the QC duality used in all these works were rather indirect.

The main goal of this paper is to give a precise formulation
of the QC duality for a rather representative class of models
together with {\it direct} proofs.
The latter require
an elaborate algebraic analysis.
We also present the brane counterparts of all its aspects
using the brane interpretation of the
duality as the relation between the 3d quiver gauge SUSY theory and
4d SUSY gauge theory with nontrivial boundary conditions  \cite{gk}.

To be precise, we give a direct proof of the following
correspondence between quantum and classical integrable systems
(Theorem 1 in Section 4).

{\em On the quantum side, consider the inhomogeneous $GL(n)$-based
generalized spin chain
of $XXX$ type with a formal Planck's constant $\hbar$ on $N$ sites with
inhomogeneity parameters $q_i$ and vector representations at each
site. Let us impose twisted boundary conditions with the twist
matrix $V=\mbox{diag} \, (V_1, V_2 , \ldots , V_n)$, with the
generating function of commuting integrals of motion (the transfer
matrix) depending on the spectral parameter $z$ being of the form
$$
T^{\hbox{\tiny{XXX}}}(z)=
\mbox{tr} \, V +\sum_{j=1}^{N}\frac{H^{\hbox{\tiny{XXX}}}_j}{z-q_j}\,.
$$
The residues $H^{\hbox{\tiny{XXX}}}_j$ are
(non-local) Hamiltonians of the
spin chain. Their eigenvalues depend on the set $\{q_i\}_N$ and
on a solution
$\Bigl \{\{\mu^1_i\}_{N_1}, \ldots , \{\mu^{n-1}_i\}_{N_{n-1}}\Bigr \}$
of the system of (nested) Bethe equations (BE):
$H^{\hbox{\tiny{XXX}}}_j=H^{\hbox{\tiny{XXX}}}_j(\{q_i\}_N;
\{\mu^1_i\}_{N_1}, \ldots , \{\mu^{n-1}_i\}_{N_{n-1}})$,
where $N_{a}$ denotes the number of Bethe roots at the
$a$-th level of the nested Bethe ansatz.

On the classical side,
consider the RS
model with coupling constant
$\hbar$ and
the number of particles, $N$, equal to the number of sites of the
$GL(n)$ spin chain. The Lax
matrix of the model is
 \beq\label{laxRS101}
 L^{\hbox{\tiny{RS}}}_{ij}(\{\dot{q}_i\}_N,
 \{q_i\}_N,\hbar )=\frac{\hbar
 \, {\dot q}_j}{q_i-q_j+\hbar}\,,\ \ \ i\,,j=1\,,...\,,N
 \eeq
 where $\{q_i\}_N$ are coordinates of the particles and
 $\{\dot q_i\}_N$ are their velocities.

The claim is that under the substitution
 \beq\label{gc5018}
{\dot q}_j=\frac{1}{\hbar}
H^{\hbox{\tiny{XXX}}}_j \Bigl (\{q_i\}_N;
\{\mu^1_i\}_{N_1}, \ldots , \{\mu^{n-1}_i\}_{N_{n-1}}\Bigr )\,,\
\ \ j=1\,,...\,,N\,,
  \eeq
  where the set of $\mu_i^a$'s is any
  solution of the nested
  BE for the spin chain, the eigenvalues of the Lax matrix
  are
 \beq\label{gc5026}
 \begin{array}{c}
 \big(\underbrace{V_1\,,\ldots\,,V_1}_{N-N_1}\,,\underbrace{V_2\,,\ldots\,,V_2}_{N_1-N_2}\,,\ldots\,,
 \underbrace{V_{n\!-\!1}\,,\ldots\,,V_{n\!-\!1}}_{N_{n\!-\!2}-N_{n\!-\!1}}\,,
 \underbrace{V_n\,,\ldots\,,V_n}_{N_{n\!-\!1}}\big ).
\end{array}
  \eeq
This means that the spectral problem
for the quantum spin chain is equivalent to an ``inverse spectral
problem'' for the Lax matrix of the classical RS system:
for the matrix of the form (\ref{laxRS101})
find velocities $\dot q_i$ in such a way that the spectrum has
the form (\ref{gc5026}).}

\noindent
    The simplest example is given in
    Section 4.

The paper is organized as follows. In Section 2 the main properties
of the classical many-body systems are
summarized. In Section 3 we review the relevant
facts concerning the quantum spin chains and in Section 4
the algebraic analysis yielding the precise correspondence
between the data at the classical and quantum sides
is presented. The brane picture behind the correspondence
considered can be found in Section 5.
A partial list of open problems
is given in the last section.

\paragraph{{\small  Acknowledgments.}} {\small The authors are grateful to
A. Alexandrov, K. Bulycheva, E. Gorsky, S. Gukov, V. Kazakov, P.
Koroteev, I. Krichever, S. Leurent, A. Morozov, N. Nekrasov, N.
Slavnov, T. Takebe and Z. Tsuboi
for discussions. The work of A.G.  was supported in part
by grants RFBR-12-02-00284 and PICS-12-02-91052. A.G. thanks the
organizers of  Simons Summer School at Simons Center for Geometry
and Physics where the part of this work has been done for the
hospitality and support. The work of A.Zabrodin was supported in part
by RFBR grant 11-02-01220, by joint RFBR grants 12-02-91052-CNRS,
12-02-92108-JSPS and by Ministry of Science and Education of Russian
Federation under contract 8207 and by grant NSh-3349.2012.2 for
support of leading scientific schools. The work of A.Zotov was
supported in part by RFBR grants 14-01-00860 and 12-02-00594, by
grant NSh-4724.2014.2 for support of leading scientific schools and
by the D. Zimin's fund "Dynasty".}

\bigskip

\section{Classical integrable many-body systems}
\setcounter{equation}{0}

\paragraph{The Ruijsenaars-Schneider (RS) model}
\cite{Ruijs1} of
${\mathfrak{gl}}_N$ type is defined by the following $N\times N$ Lax matrix
 \beq\label{gc10011}
 L^{\hbox{\tiny{RS}}}_{ij}={{\eta\nu\, e^{\eta p_j}}
 \over{q_i-q_j+\eta\nu}}\prod\limits_{k\neq
j}^N\frac{q_j-q_k+\eta\nu}{q_j-q_k}\,,\ \ \ i,j=1\,,...\,,N\,,
  \eeq
where $p_i$ and $q_i$ are the canonical variables with the Poisson
brackets
$\{p_i,q_j\}=\delta_{ij}$, $\nu$ is the coupling constant and $\eta$
is the inverse of the light speed.
Note that there is a freedom in definition (\ref{gc10011}) coming from the canonical transformation
 \beq\label{gc10012}
e^{\eta p_j}\ \longrightarrow\ e^{\eta p_j}
\prod\limits_{k\neq j}\left(\frac{q_j-q_k+\xi}{q_j-q_k-\xi}\right)^g\,,
  \eeq
where $g$ and $\xi$ are arbitrary constants. The conventional form
of the RS Lax matrix \cite{Ruijs1} is reproduced by
choosing $\xi=\pm\eta\nu$, $g=\mp\frac{1}{2}$.

The Hamiltonian of the model is
 \beq\label{gc10013}
 H^{\hbox{\tiny{RS}}}=\tr L^{\hbox{\tiny{RS}}}=
 \sum\limits_{j=1}^N e^{\eta p_j}\prod\limits_{k\neq
j}^N\frac{q_j-q_k+\eta\nu}{q_j-q_k}\,.
  \eeq
  The higher Hamiltonians in involution are
 $H^{\hbox{\tiny{RS}}}_k=\frac{1}{k}\, \tr (L^{\hbox{\tiny{RS}}})^k$,
  $H^{\hbox{\tiny{RS}}}_1= H^{\hbox{\tiny{RS}}}$.

As is seen from (\ref{gc10013}), the velocities are given by
 \beq\label{gc10014}
{\dot q}_j ={\frac{\p H}{\p p_i}}^{\hbox{\tiny{RS}}}= \eta e^{\eta p_j}\prod\limits_{k\neq
j}^N\frac{q_j-q_k+\eta\nu}{q_j-q_k}\,.
  \eeq
In terms of velocities, the Lax matrix (\ref{gc10011}) takes the form
 \beq\label{gc10015}
 L^{\hbox{\tiny{RS}}}_{ij}=\frac{\nu\, {\dot q}_j}{q_i-q_j+\eta\nu}\,,\ \ \ i,j=1\,,...\,,N.
  \eeq
The equations of motion are:
\beq\label{eqmRS}
\ddot q_i =-\sum_{k\neq i}
\frac{2\eta^2 \nu^2 \dot q_i \dot q_k}{(q_i-q_k)\bigl ((q_i-q_k)^2 \! -\!
\eta^2 \nu^2\bigr )}, \quad i=1, \ldots , N.
\eeq
In what follows we put $\eta=1$ since it can be easily restored.
But before that let us consider the non-relativistic
limit $\eta\rightarrow 0$.


\paragraph{The Calogero-Moser
(CM) model} \cite{Calogero}
is defined by the Lax matrix
 \beq\label{gc410}
L^{\hbox{\tiny{CM}}}_{ij}=
\lim\limits_{\eta\rightarrow 0}
\frac{L^{\hbox{\tiny{RS}}}_{ij}-\delta_{ij}}{\eta}
=\delta_{ij}\Bigl (p_i+{\nu}
\sum\limits_{k\neq i}\frac{1}{q_i-q_k}\Bigr )+
\nu\frac{1-\delta_{ij}}{q_i-q_j}\,,\ \ \ i,j=1\,,...\,,N\,.
  \eeq
The $\eta$-expansion of the Hamiltonian $H^{\hbox{\tiny{RS}}}$ is
$
H^{\hbox{\tiny{RS}}}=1+\eta
P^{\hbox{\tiny{CM}}}+\eta^2 H^{\hbox{\tiny{CM}}}+O(\eta^3)
$,
where
\beq\label{momentumCal}
P^{\hbox{\tiny{CM}}}=\sum_{j=1}^N
\Bigl (p_j+\sum\limits_{k\neq j}\frac{\nu}{q_j-q_k}\Bigr )=
\sum_{j=1}^N p_j\, ,
\eeq
\beq\label{gc411}
H^{\hbox{\tiny{CM}}}=\frac{1}{2}\,
\tr \left(L^{\hbox{\tiny{CM}}}\right)^2=
\frac{1}{2}\sum\limits_{i=1}^N
\Bigl (p_i+\sum\limits_{k\neq i}\frac{\nu}{q_i-q_k}\Bigr )^2-
\sum\limits_{i<j}^N \frac{\nu^2}{(q_i-q_j)^2}
  \eeq
are respectively
the total momentum and the Hamiltonian of the CM
particles.
Similarly to the $\eta \neq 0$ case, there is a freedom
to make a canonical transformation of the form
 \beq\label{gc4101}
p_j\ \rightarrow\ p_j+\nu'\sum\limits_{k\neq j}\frac{1}{q_j-q_k}
  \eeq
in (\ref{gc410}), where $\nu'$ is an arbitrary constant.
The conventional form of the CM Lax matrix
corresponds to the choice $\nu'=-\nu$.
The higher Hamiltonians in involution are
 $H^{\hbox{\tiny{CM}}}_k=\frac{1}{k}\, \tr (L^{\hbox{\tiny{CM}}})^k$, with
  $H^{\hbox{\tiny{CM}}}_1= P^{\hbox{\tiny{CM}}}$,
  $H^{\hbox{\tiny{CM}}}_2= H^{\hbox{\tiny{CM}}}$.

The particles velocities are
 \beq\label{gc412}
{\dot q}_i ={\frac{\p H}{\p p_i}}^{\hbox{\tiny{CM}}}=p_i+\sum\limits_{k\neq i}\frac{\nu}{q_i-q_k}\,.
  \eeq
In terms of the velocities, the Lax matrix and the
equations of motion acquire their conventional form:
\beq\label{gc413}
L^{\hbox{\tiny{CM}}}_{ij}=
\delta_{ij}{\dot q}_i+\nu
\frac{1-\delta_{ij}}{q_i-q_j}\,,\ \ \ i,j=1\,,...\,,N\,,
  \eeq
\beq\label{gc4137} \ddot q_i =-\sum_{k\neq i}\frac{2\,
\nu^2}{(q_i-q_k)^3}\,, \quad i=1\,,...\,,N\,.
\eeq


\section{Quantum spin chains and Gaudin models}
\setcounter{equation}{0}

\paragraph{The generalized $GL(n)$-invariant
inhomogeneous $XXX$ spin
chain \cite{KR}.}
The Hilbert space ${\cal H}$ of the model
is the tensor product of the highest weight
${\mathfrak{gl}}_n$-modules ${\mathcal M}_1\otimes ...
\otimes{\mathcal M}_N$ with the highest weights
$\la^{(1)}\,,...\,,\la^{(N)}$, $\la^{(i)}=(\la^{(i)}_1\,,...\,,\la^{(i)}_n)$
with $\la^{(i)}_1 \geq \la^{(i)}_2\geq \ldots \geq \la^{(i)}_n\geq 0$.
Let $\Lambda =\{ \la^{(1)}\,,...\,,\la^{(N)}\}$ be the set of the
highest weights. By ${\cal M}_0= {\mathbb C} ^n$
we denote the auxiliary space of the vector $GL(n)$-representation
with the highest weight $\lambda^{(0)}=(1, 0,\ldots , 0)$.

The $GL(n)$-invariant $R$-matrix $R_{0j}(z)$ acts
  non-trivially in ${\cal M}_0\otimes {\cal M}_j$.
  It has the form
 \beq\label{gc2091}
 R_{0j}(z)=1\otimes 1+\frac{\hbar}{z}
 \sum\limits_{a,b=1}^n E^{(0)}_{ab} \otimes {\sf E}^{(j)}_{ba}\,,
  \eeq
where $(E^{(0)}_{ab})_{cd}=\delta_{ac}\delta_{bd}$ are basic matrices
in the auxiliary space and ${\sf E}^{(j)}_{ab}$ are generators of
${\mathfrak{gl}}_n$ acting in ${\cal M}_j$ with the standard
commutation relations $[{\sf E}^{(j)}_{ab}, {\sf E}^{(j)}_{a'b'}]=
\delta_{a'b}{\sf E}^{(j)}_{ab'}-\delta_{ab'}{\sf E}^{(j)}_{a'b}$.

The quantum transfer matrix is an operator in ${\cal H}$ defined as trace of
a product of the $R$-matrices and the twist matrix
$V=\hbox{diag}(V_1\,,...\,,V_n)\in GL(n)$
taken in
the auxiliary space:
 \beq\label{gc2090}
 \hat T^{\hbox{\tiny{XXX}}}_{\Lambda}(z)=
 \mbox{tr}_0\, \Bigl [V_0\,R_{01}(z-q_1)\,...\,R_{0N}(z-q_N)\Bigr ]\,,
  \eeq
  where we write $V_0$ instead of $V$
  to stress that this matrix acts in ${\cal M}_0$.
  Sometimes we will use the more detailed notation
  $\hat T^{\hbox{\tiny{XXX}}}_{\Lambda}(z)=
  \hat T^{\hbox{\tiny{XXX}}}_{\Lambda}(z; \{q_i\}, V, \hbar )$.
  Hereafter we assume that the inhomogeneity parameters $q_i$ are all
  distinct.
  The Yang-Baxter equation satisfied by the $R$-matrix and the
  $GL(n)$-invariance of the $R$-matrix imply that the transfer matrices
  $\hat T^{\hbox{\tiny{XXX}}}_{\Lambda}(z)$ with the same
  $\{q_i\}$, $\Lambda$,
  $\hbar$ and $V$ commute for all values of $z$.
  Therefore, the transfer matrix can serve as a generating function for
  commuting quantum Hamiltonians. It is clear from (\ref{gc2090}) that
  $\hat T^{\hbox{\tiny{XXX}}}_{\Lambda}(z)$ has simple poles at $z=q_i$.
  The Hamiltonians can be defined as residues at these poles:
  \beq\label{Hres}
  \hat H_{\Lambda , \,i}^{\hbox{\tiny{XXX}}}:=
  \res\limits_{z=q_i}\hat T^{\hbox{\tiny{XXX}}}_{\Lambda}(z).
  \eeq
  In general, they are non-local operators involving
  spins at all sites of the chain.

As is
shown in \cite{KR} (see also \cite{MTV02,BR08}),
the eigenvalues of the transfer matrix and the Hamiltonians are of the
form
 \beq\label{gc2012}
T^{\hbox{\tiny{XXX}}}_{\Lambda}(z)=\sum\limits_{b=1}^n V_b
\, \prod\limits_{k=1}^N \frac{z-q_k+\hbar \la^{(k)}_b}{z-q_k}
\prod\limits_{\ga=1}^{N_{b\!-\!1}}\frac{z-\mu_\ga^{b\!-\!1} +\hbar}{z-\mu_\ga^{b\!-\!1}}
\prod\limits_{\ga=1}^{\,N_{b}}\frac{z-\mu_\ga^{b} -\hbar}{z-\mu_\ga^{b}}\,,
  \eeq
 \beq\label{gc20123}
\frac{1}{\hbar}H_{\Lambda , \, i}^{\hbox{\tiny{XXX}}}
= \sum\limits_{b=1}^n\, V_b \la^{(i)}_b \prod\limits_{k\neq i}^N
\frac{q_i-q_k+\hbar \la^{(k)}_b}{q_i-q_k}
\prod\limits_{\ga=1}^{N_{b\!-\!1}}\frac{q_i-\mu_\ga^{b\!-\!1} +\hbar}{q_i-\mu_\ga^{b\!-\!1}}
\prod\limits_{\ga=1}^{\,N_{b}}\frac{q_i-\mu_\ga^{b} -\hbar}{q_i-\mu_\ga^{b}},
  \eeq
where the parameters $\mu_{\gamma}^b$
satisfy the system of the nested BE:
 \beq\label{gc2013}
  \begin{array}{|c|}
  \hline\\
  \displaystyle{
{V_b}\prod\limits_{k=1}^N \frac{\!\mu^b_\be-q_k+\hbar
\la^{(k)}_b}{\,\,\mu^b_\be-q_k+\hbar\la^{(k)}_{b\!+\!1}}
 \prod\limits_{\ga=1}^{N_{b\!-\!1}}\frac{\mu^b_\be-\mu_\ga^{b\!-\!1}+
 \hbar}{\mu^b_\be-\mu_\ga^{b\!-\!1} }
=\! {V_{b\!+\!1}}\prod\limits_{\ga\neq \be}^{N_{b}}
\frac{\mu^b_\be-\mu_\ga^{b}+\hbar }{\mu^b_\be-\mu_\ga^{b}-\hbar}
\prod\limits_{\ga=1}^{N_{b\!+\!1}}\frac{\mu^b_\be-\mu_\ga^{b\!+\!1}-
\hbar }{\mu^b_\be-\mu_\ga^{b\!+\!1}}.}
\\ \ \\
\hline
\end{array}
  \eeq
Here $b\!=\!1\,,...\,,n\!-\!1$, $\be\!=\!1\,,...\,,N_b$. It is
convenient to put
$N_0=N_{n} = 0$. The total number of equations equals
$\sum\limits_{b=1}^{n-1}N_b$.
We also have \cite{KR,BR08}:
 $N\geq N_1\geq N_2\geq\ldots\geq N_{n-1}\geq 0$.

It is known \cite{BR08} that the operators
\beq\label{hatM}
\hat M_a =\sum_{j=1}^N {\sf E}_{aa}^{(j)}\,, \qquad
a=1, \ldots , n,
\eeq
commute with the transfer matrix. The eigenvectors of the latter,
built from solutions to the BE with the numbers of Bethe roots
at level $b$ equal to $N_b$, are also eigenvectors of
the operators $\hat M_a$ with the eigenvalues
\beq\label{hatM1}
M_a= N_{a-1}-N_a + \sum_{j=1}^{N} \lambda_{a}^{(j)}.
\eeq

  In what follows we consider the important particular case
  of vector representations of $GL(n)$ at all sites of the chain, i.e.,
 \beq\label{gc2014}
\la^{(i)}=(1,0\,,...\,,0)\,,\ \ \hbox{for}\ \hbox{all}\ i=1,...,N\,.
  \eeq
  In this case we will simply write $\hat T^{\hbox{\tiny{XXX}}}(z)$ and
  $\hat H^{\hbox{\tiny{XXX}}}_i$ for the transfer matrix and Hamiltonians.
Then all terms in (\ref{gc20123}) vanish except the first one:
 \beq\label{gc2015}
 \begin{array}{|c|}
  \hline\\ \displaystyle{
\frac{1}{\hbar}H^{\hbox{\tiny{XXX}}}_i=
V_1 \prod\limits_{k=1}^N \frac{q_i-q_k+\hbar}{q_i-q_k}
\prod\limits_{\ga=1}^{\,N_{1}}\frac{q_i-\mu_\ga^{1} -
\hbar}{q_i-\mu_\ga^{1}}}\,.
\\ \ \\
\hline
\end{array}
  \eeq
  The eigenvalues of the operators $\hat M_a$ are:
  $M_1=N-N_1$, $M_a=N_{a-1}-N_a$, $a=2, \ldots , n$.
The BE simplify
as well because the first product in the l.h.s. of (\ref{gc2013}) is non-trivial
only at $b=1$. The BE (\ref{gc2013}) are naturally divided into
$n-1$ groups:
 \beq\label{gc2016}
  BE_{1}: \ \
{V_1}\prod\limits_{k=1}^N \frac{\!\mu^1_\be-q_k+\hbar}{\,\,\mu^1_\be-q_k}
={V_{2}}\prod\limits_{\ga\neq \be}^{N_{1}}\frac{\mu^1_\be-\mu_\ga^{1}+
\hbar }{\mu^1_\be-\mu_\ga^{1}-\hbar}
\prod\limits_{\ga=1}^{N_{2}}
\frac{\mu^1_\be-\mu_\ga^{2}-\hbar }{\mu^1_\be-\mu_\ga^{2}}\,,
  \eeq
 \beq\label{gc2018}
BE_{\, b}:\ \
{V_b}
 \prod\limits_{\ga=1}^{N_{b\!-\!1}}\frac{\mu^b_\be-\mu_\ga^{b\!-\!1}+
 \hbar}{\mu^b_\be-\mu_\ga^{b\!-\!1} }
={V_{b\!+\!1}}\prod\limits_{\ga\neq \be}^{N_{b}}
\frac{\mu^b_\be-\mu_\ga^{b}+\hbar }{\mu^b_\be-\mu_\ga^{b}-\hbar}
\prod\limits_{\ga=1}^{N_{b\!+\!1}}
\frac{\mu^b_\be-\mu_\ga^{b\!+\!1}-\hbar}{\mu^b_\be-\mu_\ga^{b\!+\!1}}
  \eeq
for $b=2\,,...,\,n\!-\!2$ and
 \beq\label{gc2019}
  BE_{\, n\! -\! 1}:\ \
{V_{n-1}}
\prod\limits_{\ga=1}^{N_{n\!-\!2}}
\frac{\mu^{n\!-\!1}_\be-\mu_\ga^{n\!-\!2}+\hbar}{\mu^{n\!-\!1}_\be-
\mu_\ga^{n\!-\!2} }
={V_{n}}\prod\limits_{\ga\neq \be}^{N_{n\!-\!1}}\frac{\mu^{n\!-\!1}_\be-\mu_\ga^{n\!-\!1}+\hbar
}{\mu^{n\!-\!1}_\be-\mu_\ga^{n\!-\!1}-\hbar}\,.
  \eeq
In what follows we will use the notation
$H^{\hbox{\tiny{XXX}}}_i (\{q_i\}_N , \{ \mu_{\alpha}^1\}_{N_1})$
for the function given by the r.h.s. of (\ref{gc2015}). When
the set $\{ \mu_{\alpha}^1\}_{N_1}$ is taken from a solution
to the system of BE, this function is equal to an eigenvalue
of the Hamiltonian.

\vskip2mm

\noindent {\em \underline{Example:} $GL(2)$ XXX chain}

\noindent  In this case the twist matrix is
$V=\mat{\,e^\om}{0}{0}{e^{-\om}}$ and eigenvalues of
the transfer matrix and the Hamiltonians are given by
 \beq\label{gc302}
 T^{\hbox{\tiny{XXX}}}_{\Lambda}(z)
 =e^{\om}\prod\limits_{k=1}^N\frac{z-q_k+
 \hbar\la_1^{(k)}}{z-q_k}\prod\limits_{\ga=1}^{N_1}
 \frac{z-\mu_\ga-\hbar}{z-\mu_\ga}+
 e^{-\om}\prod\limits_{k=1}^N
 \frac{z-q_k+\hbar\la_2^{(k)}}{z-q_k}\prod
 \limits_{\ga=1}^{N_1}\frac{z-\mu_\ga+\hbar}{z-\mu_\ga}
  \eeq
$$
\frac{1}{\hbar}H^{\hbox{\tiny{XXX}}}_{\Lambda ,\,i}
=
 $$
 \beq\label{gc303}
e^{\om}\la^{(i)}_1\prod \limits_{k\neq i}^N\frac{q_i\! -\! q_k \!
+\! \hbar\la^{(k)}_1}{q_i-q_k}\prod\limits_{\ga=1}^{N_1} \frac{q_i
\! -\! \mu_\ga \! -\! \hbar}{q_i-\mu_\ga}+
 e^{-\om}\la^{(i)}_2\prod\limits_{k\neq i}^N\frac{q_i\! -\! q_k\! +\! \hbar\la^{(k)}_2}{q_i-q_k}\prod\limits_{\ga=1}^{N_1}
 \frac{q_i\! -\! \mu_\ga \! +\! \hbar}{q_i-\mu_\ga}
\eeq
with the BE of the form
 \beq\label{gc304}
e^{2\om}\prod\limits_{k=1}^N
\frac{\mu_\al-q_k+\hbar\la_1^{(k)}}{\mu_\al-q_k+\hbar\la_2^{(k)}}
=\prod\limits_{\ga\neq\al}^{N_1}
\frac{\mu_\al-\mu_\ga+\hbar}{\mu_\al-\mu_\ga-\hbar}\,,\qquad \al=1\,,...\,,N_1\,.
  \eeq
With the choice $(\la_1^{(i)},\la_2^{(i)})=(1,0)$
for all $i=1,\ldots ,N$ (spin $\frac{1}{2}$ at
each site) the second term in (\ref{gc303})
vanishes and we get
 \beq\label{gc305}
\frac{1}{\hbar}H^{\hbox{\tiny{XXX}}}_i
=e^{\om}\prod\limits_{k\neq
i}^N\frac{q_i-q_k+\hbar}{q_i-q_k}\prod
\limits_{\ga=1}^{N_1}\frac{q_i-\mu_\ga-\hbar}{q_i-
\mu_\ga}\,,\qquad i=1\,,\ldots\,,N,
  \eeq
where $\mu_{\alpha}$'s satisfy the BE
 \beq\label{gc306}
e^{2\om}\prod\limits_{k=1}^N\frac{\mu_\al-q_k+\hbar}{\mu_\al-q_k}
=\prod\limits_{\ga\neq\al}^{N_1}
\frac{\mu_\al-\mu_\ga+\hbar}{\mu_\al-\mu_\ga-\hbar}\,,\qquad
\al=1\,,...\,,N_1\,.
  \eeq

\paragraph{The rational ${\mathfrak{gl}}_n$ Gaudin model} \cite{Gaudin}
is the $\varepsilon \to 0$ limit of the inhomogeneous $XXX$ spin
chain with the transfer matrix $\hat
T^{\hbox{\tiny{XXX}}}_{\Lambda}(z; \{q_i\}, V^{\varepsilon},
\varepsilon \hbar )$. The expansion as $\varepsilon \to 0$ is
 \beq\label{expansion} \hat T^{\hbox{\tiny{XXX}}}_{\Lambda}(z;
\{q_i\}, V^{\varepsilon}, \varepsilon \hbar )= n+\varepsilon \left (
\mbox{tr}\, v + \sum_{i=1}^N \frac{\hbar \, C_1^{(i)}}{z-q_i}\right
) + \varepsilon^2  \left (\frac{1}{2}\, \mbox{tr}\, v^2
+\sum_{i}\frac{\hbar \, \hat H^{\hbox{\tiny{G}}}_{\Lambda , \,
i}}{z-q_i} \right ) +O(\varepsilon ^3)
 \eeq where $v=\log V$,
$\displaystyle{ C_1^{(i)}=\sum_a {\sf E}_{aa}^{(i)}
=\sum_{a}\lambda_a^{(i)}}$ is the first Casimir operator of
$U({\mathfrak{gl}}_n)$ at the $i$-th site and
 \beq\label{HG} \hat
H^{\hbox{\tiny{G}}}_{\Lambda , \, i}= \sum_a v_a {\sf E}_{aa}^{(i)}+
\sum_{j\neq i} \frac{\hbar}{q_i-q_j} \sum_{ab} {\sf
E}_{ab}^{(i)}{\sf E}_{ba}^{(j)} =\lim\limits_{\varepsilon \to 0}
\frac{\hat H^{\hbox{\tiny{XXX}}}_{\Lambda , \, i}(\{q_i\},
e^{\varepsilon v}, \varepsilon \hbar ) -\varepsilon \hbar
C_1^{(i)}}{\hbar \varepsilon^2}\,.
 \eeq
These operators are called
Gaudin Hamiltonians. Their eigenvalues can be found by substituting
(\ref{gc20123}) into (\ref{HG}) and tending $\varepsilon\rightarrow
0$. This gives
 \beq\label{gc203}
H_{\Lambda , \,i}^{\hbox{\tiny{G}}}=\sum\limits_{b=1}^n\, \Big ( v_b+\sum\limits_{k\neq i}^N \frac{\hbar\la^{(k)}_b\la^{(i)}_b}{q_i-q_k}
+\sum\limits_{\ga=1}^{N_{b\!-\!1}}\frac{\hbar\la^{(i)}_b}{q_i-\mu_\ga^{b\!-\!1}}
-\sum\limits_{\ga=1}^{\,N_{b}}\frac{\hbar\la^{(i)}_b}{q_i-\mu_\ga^{b}}\Big)
  \eeq
with the BE of the form
 \beq\label{gc204}
{v_b}-{v_{b\!+\!1}}+\sum\limits_{k=1}^N \frac{\hbar(\la^{(k)}_b-
\la^{(k)}_{b\!+\!1})}{\,\,\mu^b_\be-q_k}=
 -\sum\limits_{\ga=1}^{N_{b\!-\!1}}\frac{\hbar}{\mu^b_\be-\mu_\ga^{b\!-\!1}}
 +2\sum\limits_{\ga\neq \be}^{N_{b}}\frac{\hbar }{\mu^b_\be-\mu_\ga^{b}}
-\sum\limits_{\ga=1}^{N_{b\!+\!1}}\frac{\hbar }{\mu^b_\be-\mu_\ga^{b\!+\!1}}\,,
  \eeq
where $b\!=\!1\,,...\,,n\!-\!1$, $\be\!=\!1\,,...\,,N_b$
$N_0 \! =\! N_{n}\!=\!0$. By analogy with the $XXX$ spin chain we call
the matrix $v=\mbox{diag}(v_1\,,...\,,v_n)$ the twist matrix of the
Gaudin model. In the context of the Gaudin model, the
parameters $q_i$ are often called marked points.

The operators $\hat M_a$ and their eigenvalues
on the eigenstates of the Gaudin Hamiltonians are
are given by the same formulas (\ref{hatM}), (\ref{hatM1}).
In the case $\lambda^{(i)} =(1, 0, \ldots , 0)$ for all $i$ the above
formulas simplify:
 \beq\label{gc205}
  \begin{array}{|c|}
  \hline\\ \displaystyle{
H_i^{\hbox{\tiny{G}}}=v_1+\sum\limits_{k\neq i}^N \frac{\hbar}{q_i-q_k}
%
%
-\sum\limits_{\ga=1}^{N_{1}}\frac{\hbar}{q_i-\mu_\ga^{1}}}
\\ \ \\
\hline
\end{array}
  \eeq
with the $BE_{b}:$
 \beq\label{gc206}
  \begin{array}{|c|}
  \hline\\
 \displaystyle{
{v_b}-{v_{b\!+\!1}}+\delta_{1b}\sum\limits_{k=1}^N
\frac{\hbar}{\,\,\mu^b_\be-q_k}=
 -\sum\limits_{\ga=1}^{N_{b\!-\!1}}\frac{\hbar}{\mu^b_\be-\mu_\ga^{b\!-\!1}}
 +2\sum\limits_{\ga\neq \be}^{N_{b}}\frac{\hbar }{\mu^b_\be-\mu_\ga^{b}}
-\sum\limits_{\ga=1}^{N_{b\!+\!1}}\frac{\hbar
}{\mu^b_\be-\mu_\ga^{b\!+\!1}}}
\\ \ \\
\hline
\end{array}
  \eeq
  The eigenvalues of the operators $\hat M_a$ are:
  $M_1=N-N_1$, $M_a=N_{a-1}-N_a$, $a=2, \ldots , n$.
Similarly to the $XXX$ spin chain case, we will use the notation
$H^{\hbox{\tiny{G}}}_i (\{q_i\}_N , \{ \mu_{\alpha}^1\}_{N_1})$
for the function given by the r.h.s. of (\ref{gc205}). When
the set $\{ \mu_{\alpha}^1\}_{N_1}$ is taken from a solution
to the system of BE (\ref{gc204}), this function is equal to an eigenvalue
of the Hamiltonian.

\vspace{2mm}

\noindent {\em \underline{Example:}
The rational ${\mathfrak{gl}}_2$ Gaudin model}
 \vskip2mm
\noindent For the ${\mathfrak{gl}}_2$ Gaudin model with
the twist matrix
 $
v=\mat{\om}{0}{0}{-\om}
  $
equations (\ref{gc205}) and (\ref{gc206}) read
 \beq\label{gc371}
\frac{1}{\hbar}H_i^{\hbox{\tiny{G}}}=
\om+\sum\limits_{k\neq i}^N\frac{\hbar}{q_i-q_k}+
\sum\limits_{\ga=1}^{N_1}\frac{\hbar}{\mu_\ga-q_i}\,,
  \eeq
 \beq\label{gc372}
2\om+ \hbar\sum\limits_{k=1}^N\frac{1}{\mu_\al-q_k}
=2\hbar\sum\limits_{\ga\neq\al}^{N_1} \frac{1}{\mu_\al-\mu_\ga}
  \eeq
for all $\al=1\,,...\,,{M}$. Here $\mu_\al=\mu^1_\al$.

\section{The QC duality}
\setcounter{equation}{0}

In this section we derive the relation between the spectrum of the
quantum $XXX$ spin chain Hamiltonians and the spectrum
of the classical RS Lax matrix which is the basis of the QC duality.
Our main statement is the following theorem.

\begin{theor}\label{theor01}
Given the Lax matrix (\ref{gc10015}) of the ${\mathfrak{gl}}_N$ RS
model
 $$
 L^{\hbox{\tiny{RS}}}_{ij}(\{\dot{q}\}_N,
 \{q\}_N, \nu)=\frac{\nu\, {\dot q}_j}{q_i-q_j+\nu}\,,\ \ \ i\,,j=1\,,...\,,N,
 $$
make the substitution
 \beq\label{gc501}
 \begin{array}{|c|}
  \hline\\
  \displaystyle{
 \nu = \hbar \qquad \mbox{and} \qquad
\dot q_j=\frac{1}{\hbar} H_j^{\hbox{\tiny{XXX}}}\left (\{ q_i
\}_N\,, \{\mu _{\alpha}^{1} \}_{N_1} \right )}
\\ \ \\
\hline
\end{array}
  \eeq
where the r.h.s. is given by (\ref{gc2015}). If the $N_1$ parameters
$\mu _{\alpha}^{1} $ are taken from any solution
$\left \{ \{\mu _{\alpha}^{1} \}_{N_1},\,\ldots \,
\{\mu _{\alpha}^{n-1} \}_{N_{n-1}}\right \}$
to the system
of BE (\ref{gc2016})--(\ref{gc2019}) for the inhomogeneous spin chain
on $N\geq n$ sites with
inhomogeneity parameters $q_i$ and
the twist matrix $V=\mbox{diag} \, (V_1, \ldots , V_n)$, then
the spectrum
of the Lax matrix has the following form:
 \beq\label{gc502}
 \begin{array}{|c|}
  \hline\\
  \displaystyle{
 \hbox{Spec} \, L^{\hbox{\tiny{RS}}}
 \left (\frac{1}{\hbar}\left  \{ H_j^{\hbox{\tiny{XXX}}}\right \}_N,
 \left \{ q _j\right \} _N, \, \hbar \right )\Bigr |_{BE}}\\ \ \\
 \displaystyle{=\, \big (\underbrace{V_1\,,\ldots\,,V_1}_{N-N_1}\,, \, \underbrace{V_2\,,\ldots\,,V_2}_{N_1-N_2}\,,\ldots\,,
 \, \underbrace{V_{n\!-\!1}\,,\ldots\,,
 V_{n\!-\!1}}_{N_{n\!-\!2}-N_{n\!-\!1}}\,,\,
 \underbrace{V_n\,,\ldots\,,V_n}_{N_{n\!-\!1}}\big)}
\\ \ \\
\hline
\end{array}
  \eeq
\end{theor}

The rest of this section is devoted to the proof of the
theorem. But before passing to the proof, let us say a few
words about the meaning of this statement. It implies,
in particular, that one can solve
the spectral problem for the Hamiltonians of the
inhomogeneous spin chain
without addressing the BE at any step but by solving
an ``inverse spectral problem''
for the Lax matrix $L^{\hbox{\tiny{RS}}}$ of the
classical RS system of particles. More precisely,
let $\{ q_i\}_N$ be the inhomogeneity parameters
of the spin chain with the Planck's constant $\hbar$ and $V$ its twist matrix.
Let the eigenvalues of the RS Lax matrix
be equal to the eigenvalues $V_a$ of the twist matrix,
with some multiplicities $M_a \geq 0$ such that
$\displaystyle{\sum_{a}M_a =N}$. (This fixes
values of all the RS Hamiltonians:
$\displaystyle{H_{k}^{\hbox{\tiny{RS}}}=\frac{1}{k}
\sum_a M_aV_a^k}$.)
Then the spectrum of $\hat H_j^{\hbox{\tiny{XXX}}}$ in the
sector where eigenvalues of the operators $\hat M_a$ are equal to
$M_a$ is given by the values of $H_j^{\hbox{\tiny{XXX}}}$ such that
the matrix
$\displaystyle{
L^{\hbox{\tiny{RS}}}_{ij}=\frac{H_j^{\hbox{\tiny{XXX}}}}{q_i-q_j+\hbar }}
$
has the prescribed spectrum.
Here we assume that
each site carries the vector representation of $GL(N)$.
We anticipate that this approach can be extended to the
spin chains with arbitrary highest weight representations
at sites.

The proof will use the results of
\cite{AASZ2}.
In order to prove the statement, i.e.,
 \beq\label{gc503}
\det  \left [
L^{\hbox{\tiny{RS}}}
 \left (\frac{1}{\hbar}\left  \{ H_j^{\hbox{\tiny{XXX}}}\right \}_N,
 \left \{ q _j\right \}_N, \, \hbar \right )\Bigr |_{BE}
-\la \right ]\, =\, \prod_{a=1}^n (V_a-\la )^{M_a},
  \eeq
where $M_1=N\! -\! N_1$, $M_a = N_{a-1}\! -\! N_a$ ($2\leq a\leq n$)
 let us introduce the following pair of matrices:
 \beq\label{gc504}
 {\mathcal{L}}_{ij}(\{x_i\}_N,\{y_i \}_M,g)=
 \frac{g\,\hbar}{x_i-x_j+\hbar}\prod\limits_{k\neq j}^N\frac{x_j-x_k+\hbar}{x_j-x_k}
\prod\limits_{\ga=1}^M\frac{x_j-y_\ga}{x_j-y_\ga+\hbar}\,,
\ \ \ i\,,j=1\,,...\,,N
  \eeq
and
 \beq\label{gc505}
{\widetilde {\mathcal{L}}}_{\al\be}(\{y_i \}_M,\{x_i\}_N,g)=
\frac{g\,\hbar}{y_\al-y_\be+\hbar}\prod\limits_{\ga\neq
\be}^M\frac{y_\be-y_\ga-\hbar}{y_\be-y_\ga}
\prod\limits_{k=1}^N\frac{y_\be-x_k}{y_\be-x_k-\hbar}\,,\ \
\al,\be=1\,,...\,,M\,,
  \eeq
From the computational point of view
the QC duality is based on the following algebraic
relation between ${\mathcal{L}}$ and ${\widetilde
{\mathcal{L}}}$:
 \begin{predl}\label{predl01}
For the pair of matrices (\ref{gc504}) and (\ref{gc505}) it holds:
 \beq\label{gc506}
\begin{array}{|c|}
  \hline\\
 \det\limits_{N\times N}
 \Bigl ({\mathcal{L}}\left (\{x_i\}_N,\{y_i \}_M,g\right )
 -\la\Bigr )=(g-\la)^{N-M}
\det\limits_{M\times M}\Bigl ({\widetilde {\mathcal{L}}}
\left (\{y_i \}_M,\{x_i\}_N,g \right )-\la\Bigr )\,
\\ \ \\
\hline
\end{array}
  \eeq
 \end{predl}
The proof of (\ref{gc506}) is given in the Appendix.

\vskip3mm\underline{\em{Proof of Theorem \ref{theor01}}}:\vskip3mm

\noindent The proof of (\ref{gc503})
includes $n-1$ steps and consists in successive application
of (\ref{gc506}) and taking into account the BE (\ref{gc2016})-(\ref{gc2019}). Indeed, set
 \beq\label{gc510}
 \begin{array}{lll}
 L^{(0)}_{ij}\,\, &=& \,\, \displaystyle{L_{ij}^{\hbox{\tiny{RS}}}
 (\frac{1}{\hbar}\{H_j^{\hbox{\tiny{XXX}}}\}_N, \{ q_i\}_N,\hbar)
 =\frac{\hbar\, V_1}{q_i-q_j+\hbar}
 \prod\limits_{k\neq j}^N\frac{q_j-q_k+\hbar}{q_j-q_k}
\prod\limits_{\ga=1}^{N_1}\frac{q_j-\mu^1_\ga-\hbar}{q_j-\mu^1_\ga}}
\\ && \\
& = &{\mathcal{L}}_{ij}(\{q_i\! -\! \hbar \}_N,\{\mu_{\alpha}^{\,1}
\}_{N_1},V_1)\,,
\end{array}
  \eeq
   and define (at the first step)
 \beq\label{gc511}
 \begin{array}{l}
 L^{(1)}_{\al\be}=\displaystyle{
 {\widetilde {\mathcal{L}}}_{\al\be}(\{
 \mu_{\alpha}^{\,1}\}_{N_1},\{q_i\!-\!\hbar\}_N,V_1)=
 \frac{\hbar\, V_1}{\mu^{1}_\al-\mu^{1}_\be+\hbar}\prod\limits_{\ga\neq
\be}^{N_1}\frac{\mu^{1}_\be-\mu^{1}_\ga-\hbar}{\mu^{1}_\be-\mu^{1}_\ga}
\prod\limits_{k=1}^N\frac{\mu^{1}_\be-q_k+\hbar}{\mu^{1}_\be-q_k}\,,}
\end{array}
  \eeq
 where $\al,\be=1\,,...\,,N_1$.
Equation (\ref{gc506}) implies that
 \beq\label{gc512}
\det_{N\times N}({L^{(0)}}-\lambda)=(V_1-\la)^{N-N_1}\det_{N_1\times N_1}({L^{(1)}}-\lambda)\,.
  \eeq
Next, impose BE (\ref{gc2016}) to get:
 \beq\label{gc513}
 \left.L^{(1)}_{\al\be}\right|_{BE_1}=
  \frac{\hbar\, V_2}{\mu^{1}_\al-\mu^{1}_\be+\hbar}\prod\limits_{\ga\neq
\be}^{N_1}\frac{\mu^{1}_\be-\mu^{1}_\ga+\hbar}{\mu^{1}_\be-\mu^{1}_\ga}
\prod\limits_{\ga=1}^{N_2}\frac{\mu^{1}_\be-\mu^{2}_\ga-\hbar}{\mu^{1}_\be-\mu^{2}_\ga}\,,\qquad \al,\be=1\,,...\,,N_1\,,
  \eeq
i.e.,
 \beq\label{gc514}
 \left.L^{(1)}\right|_{BE_1}=
 {\mathcal{L}}(\{\mu_{\alpha}^{\,1}\!-\!\hbar\}_{N_1},\{
\mu_{\alpha}^{\,2}\}_{N_2},V_2)\,.
  \eeq
At the second step we define
 \beq\label{gc515}
 L^{(2)}_{\al\be}=
 {\widetilde {\mathcal{L}}}_{\al\be}(\{
\mu_{\gamma}^{\,2}\}_{N_2},\{\mu_{\gamma}^{\,1}\!-\!\hbar
\}_{N_1},V_2)\,,\qquad \al,\be=1\,,...\,,N_2\,,
  \eeq
and, similarly to the previous step,
 we use (\ref{gc506}) and BE (\ref{gc2018}) to get:
 \beq\label{gc5150}
\det_{N_1\times N_1}({L^{(1)}}-\lambda)=(V_2-\la)^{N_1-N_2}\det_{N_2\times N_2}({L^{(2)}}-\lambda)\,,
  \eeq
 \beq\label{gc5151}
 \left.L^{(2)}\right|_{BE_2}={\mathcal{L}}
 (\{\mu_{\alpha}^{\,2}\!-\!\hbar\}_{N_2},\{
\mu_{\alpha}^{\,3}\}_{N_3},V_3)\,.
  \eeq
 $$
\vdots
 $$
and so on until the last step, where we use (\ref{gc2019}):
 \beq\label{gc516}
 \left.L^{(n\!-\!1)}_{\al\be}\right|_{BE_{n\!-\!1}}=
  \frac{\hbar\, V_n}{\mu^{n\!-\!1}_\al-\mu^{n\!-\!1}_\be+\hbar}
\prod\limits_{\ga\neq\be}^{N_{n\!-\!1}}\frac{\mu^{n\!-\!1}_\be-\mu^{n\!-\!1}_\ga+\hbar}{\mu^{n\!-\!1}_\be-\mu^{n\!-\!1}_\ga}\,,\qquad
\al,\be=1\,,...\,,N_{n\!-\!1}\,.
  \eeq
The latter matrix obeys the equation
$\displaystyle{\det_{N_{n\!-\!1}\times
N_{n\!-\!1}}({L^{({n\!-\!1})}}-\lambda)=(V_n-\la)^{N_{n\!-\!1}}}$
which follows from
Proposition \ref{predl01} for $N=N_{n\!-\!1}$ and $M=0$.
$\blacksquare$

\vskip5mm

In the limiting case we have the QC duality between the
quantum Gaudin and the classical
CM models. The following analogue of Theorem \ref{theor01}
holds true.

\begin{theor}\label{theor02}
Given the Lax matrix (\ref{gc413}) of the ${\mathfrak{gl}}_N$ CM
model
 $$
L^{\hbox{\tiny{CM}}}_{ij}= \delta_{ij}{\dot q}_i+
\nu\frac{1-\delta_{ij}}{q_i-q_j}\,,\ \ \ i,j=1\,,...\,,N,
 $$
make the substitution
 \beq\label{gc5017}
 \nu = \hbar \qquad \mbox{and} \qquad
\dot q_j=\frac{1}{\hbar}
H_j^{\hbox{\tiny{G}}}\left (\{ q_i \}_N\,, \{\mu _{\alpha}^{1} \}_{N_1}
\right )\,,
\ \ j=1\,,...\,,N\,,
 \eeq
where the r.h.s. is given by (\ref{gc205}).
If the $N_1$ parameters
$\mu _{\alpha}^{1} $ are taken from any solution
$\left \{ \{\mu _{\alpha}^{1} \}_{N_1},\,\ldots \,
\{\mu _{\alpha}^{n-1} \}_{N_{n-1}}\right \}$
to the system of BE (\ref{gc206}) for the
${\mathfrak{gl}}_n$  Gaudin model with $N\geq n$ marked points
$q_i$ and the twist matrix $v=\mbox{diag} \, (v_1, \ldots , v_n)$, then
the spectrum
of the Lax matrix has the following form:
 \beq\label{gc5027}
 \begin{array}{c}
 \hbox{Spec} \, L^{\hbox{\tiny{CM}}}
 \left (\frac{1}{\hbar}\left  \{ H_j^{\hbox{\tiny{G}}}\right \}_N,
 \left \{ q _j\right \} _N, \, \hbar \right )\Bigr |_{BE}\\ \ \\
 =\, \big (\underbrace{v_1\,,\ldots\,,v_1}_{N-N_1}\,, \,
 \underbrace{v_2\,,\ldots\,,v_2}_{N_1-N_2}\,,\ldots\,,
 \, \underbrace{v_{n\!-\!1}\,,\ldots\,,
 v_{n\!-\!1}}_{N_{n\!-\!2}-N_{n\!-\!1}}\,,\, \underbrace{v_n\,,\ldots\,,v_n}_{N_{n\!-\!1}}\big)
\end{array}
  \eeq
 \end{theor}

The proof is based on the analogue of Proposition \ref{predl01}.
Introduce the pair of matrices
 \beq\label{gc017}
\begin{array}{c}
\displaystyle{ {\mathcal{L}}_{ij}(\{x_i\}_N,\{y_i\}_M,\omega)=
\delta_{ij}\left(\om+\sum\limits^N_{k\neq
i}\frac{\hbar}{q_i-q_k}+\sum\limits_{\ga=1}^M\frac{\hbar}{\mu_\ga-q_i}
\right)+(1-\delta_{ij})\frac{\hbar}{q_i-q_j}\,,}
\end{array}
  \eeq
where $i\,,j=1\,,...\,,N$ and
 \beq\label{gc032}
\begin{array}{c}
\displaystyle{{\widetilde {\mathcal{L}}}_{\al\be}
(\{y_i\}_M,\{x_i\}_N,\omega)= \delta_{\al\be}\left(
\om-\sum\limits^M_{\ga\neq\al}\frac{\hbar}{\mu_\al\!-\!\mu_\ga}
-\sum\limits^N_{k=1}\frac{\hbar}{q_k\!-\!\mu_\al}\right)+
\left(1-\delta_{\al\be}\right) \frac{\hbar}{\mu_\al\!-\!\mu_\be}\,,
}
\end{array}
  \eeq
where $\al,\be=1\,,...\,,M$.
The relation between them is given by
 \begin{predl}\label{predl02}
For the pair of matrices (\ref{gc017}) and (\ref{gc032}) it holds:
 \beq\label{gc5067}
 \det_{N\times N}
 \Bigl ({\mathcal{L}}(\{x_i\}_N,\{y_i\}_M,\om)-\la\Bigr )=(\om-\la)^{N-M}
\det_{M\times M}
\Bigl ({\widetilde {\mathcal{L}}}(\{ y_i\}_M,\{x_i\}_N,\om)-\la\Bigr )\,.
  \eeq
 \end{predl}
The proof is given in the Appendix.

\bigskip

We conclude this section by the simplest example of the
correspondence between the spectra of Gaudin Hamiltonians
and diagonal elements of the CM Lax matrix with fixed eigenvalues.

\bigskip

\noindent{\bf The simplest example}:  the rational ${\mathfrak
gl}_2$ Gaudin model with 2 marked points (sites) $q_{1,2}=\pm q$
with spins $\frac{1}{2}$ and the twist matrix
 $
v=\mat{\om}{0}{0}{-\om}
 $.
There Hilbert space of the model is 4-dimensional
and the states are classified according to
eigenvalues of
the spin $z$-projection operator:

$1)\, $ $\left(\begin{array}{l}1\\ 0\end{array}\right)$,
$\left(\begin{array}{l}1\\ 0\end{array}\right)$. In this case there
are no Bethe roots and the spectrum is given by (\ref{gc371}):
 \beq\label{gc1200}
\frac{1}{\hbar}H_{1,2}^{\hbox{\tiny{G}}}=\om\pm\frac{\hbar}{2q}\,.
  \eeq

$2)\,$ $\left(\begin{array}{l}1\\ 0\end{array}\right)$,
$\left(\begin{array}{l}0\\ 1\end{array}\right)$ or
$\left(\begin{array}{l}0\\ 1\end{array}\right)$,
$\left(\begin{array}{l}1\\ 0\end{array}\right)$. The single Bethe
root satisfies the BE
 \beq\label{gc1201}
-2\om=\frac{\hbar}{\mu-q}+\frac{\hbar}{\mu+q}\,.
  \eeq
Substituting its solution(s) $\mu_{\pm}=-\frac{\hbar}{2\om}\pm
\frac{\sqrt{4\om^2q^2+\hbar^2}}{2\om}$ we get the spectrum:
 \beq\label{gc1202}
\frac{1}{\hbar}H_{1}^{\hbox{\tiny{G}}}
\left.\right|_{\mu=\mu_\mp}=\pm\frac{\sqrt{4\om^2q^2+\hbar^2}}{2q}\,,\
\ \
\frac{1}{\hbar}H_{2}^{\hbox{\tiny{G}}}
\left.\right|_{\mu=\mu_\mp}=\mp\frac{\sqrt{4\om^2q^2+\hbar^2}}{2q}\,.
  \eeq

$3)\,$ $\left(\begin{array}{l}0\\ 1\end{array}\right)$,
$\left(\begin{array}{l}0\\ 1\end{array}\right)$. Two Bethe roots
satisfy the following BE:
 \beq\label{gc1204}
\left\{\begin{array}{l}\displaystyle{
-2\om+\frac{2\hbar}{\mu_1-\mu_2}=\frac{\hbar}{\mu_1-q}+
\frac{\hbar}{\mu_1+q}}\,,\\ \ \\ \displaystyle{
-2\om+\frac{2\hbar}{\mu_2-\mu_1}=\frac{\hbar}{\mu_2-q}+
\frac{\hbar}{\mu_2+q}\,.}
\end{array}\right.
  \eeq
The solutions $\mu_1=-\frac{\hbar}{2\om}\pm
\frac{\sqrt{4\om^2q^2-\hbar^2}}{2\om}$ and
$\mu_2=-\frac{\hbar}{2\om}\mp \frac{\sqrt{4\om^2q^2-\hbar^2}}{2\om}$
lead to the spectrum
 \beq\label{gc1205}
\frac{1}{\hbar}H_{1,2}^{\hbox{\tiny{G}}}=-\om\pm\frac{\hbar}{2q}\,.
  \eeq

Let us obtain the same spectrum from the classical rational 2-body
CM model with the coupling constant $\hbar$.
The Lax matrix is
 $$
L^{\hbox{\rm {\tiny CM}}}=\mat{{\dot
q}_1}{\frac{\hbar}{q_1-q_2}}{-\frac{\hbar}{q_1-q_2}}{{\dot q}_2}
 $$
Let us put $q_{1,2}=\pm q$ as in the Gaudin model.
The requirement for this matrix to have eigenvalues $(v_1\,,v_2)$
provides the following values of the velocities:
 \beq\label{gc1207}
{\dot q}_{1,2}=\frac{v_1+v_2}{2}
\pm\sqrt{\frac{(v_1-v_2)^2}{4}+\frac{\hbar^2}{4q^2}}\,.
  \eeq
The QC duality claims that
 ${\dot q}_{1,2}=\frac{1}{\hbar}H_{1,2}^{\hbox{\tiny{G}}}$.
The above described three cases follow from (\ref{gc1207}) when
$1)$ $v_1=v_2=\om$,
$2)$ $v_1=\om$, $v_2=-\om$,
$3)$ $v_1=v_2=-\om$.

\section{Relation to branes and gauge theories}
\setcounter{equation}{0}

In this section we briefly comment on the realization of the QC duality
in terms of branes and gauge theories.

First let us note that
there are two types of dualities in the integrable systems relevant
to our discussion: the bispectrality and the QC duality.
The bispectral transformations preserve the class of
quantum spin chains. Roughly speaking, the inhomogeneity
parameters and twists get interchanged under this
transformation.
On the classical level, the bispectrality acts by interchanging
coordinates and eigenvalues of the Lax operator and
preserves the class of
CM-RS models of different kinds (rational, trigonometric, elliptic).
The mapping can be defined for the classical and
quantum models independently with the clear semiclassical picture in
between.
One can also show that spectral curves (on the classical level)
or the systems of BE (on the quantum level) for the bispectrally dual
models are related in a
controllable way.

In distinction of the bispectrality,
the QC duality we have focused on in this paper
is a relation between representatives from the two different families
of models: quantum (spin chains, Gaudin) and classical (CM-RS).
The spin chain inhomogeneity parameters, twists and Hamiltonians
get mapped respectively to coordinates of the CM-RS particles
and eigenvalues of the Lax matrix while the Hamiltonians of
the spin chain get mapped to velocities of the CM-RS particles.
The spectral problem for Hamiltonians of the
spin chain (equivalent to solving the BE) corresponds to a
bit unusual problem at the classical CM-RS side: given values of
all integrals of motion in involution,
we should fix all coordinates and look for the
allowed values of particles velocities (or just momenta
in the CM case).
In other words, the quantum eigenstates
are encoded by intersection points of two Lagrangian
submanifolds in the phase space of a classical integrable model.

The interpretation of bispectrality in terms of the gauge
theories on the brane worldvolumes in the simplest cases has been
found at the CM-RS side in \cite{fock} and at the spin chain side in
\cite{ggm}.
More recently, the bispectrality transformation has been
used to prove the AGT duality \cite{zotov01,zz02} and the
bulk-worldsheet 2d/4d duality for the nonabelian strings \cite{bcgk}
in the integrability framework. The comprehensive analysis of the
bispectrality for the general case has been developed in \cite{gk}.
It was identified as the mirror transformation in the quiver 3d
theory with the generic matter in the fundamental representation.
Moreover, the QC duality was interpreted there
in the same quiver set-up which
encodes the particular brane configuration responsible for the gauge
theory \cite{gk}.

The brane configuration relevant to our quiver gauge theory is
as follows. We have $n$ parallel  NS5 branes  extended in the
$(x_0,x_1,x_2, x_7,x_8,x_9)$ directions, $N_i$ D3 branes extended
in $(x_0,x_1,x_2,x_3)$ between $i$-th and $(i+1)$-th NS5 branes, and
$K_i$ D5 branes extended in $(x_0,x_1,x_2,$ $x_3,x_4,x_5,x_6)$
directions  between the $i$-th and $(i+1)$-th NS5 branes. From this brane
configuration we obtain the $\prod_{i}^{n} U(N_i)$ gauge group on
the D3 brane worldvolume with additional $K_i$ fundamentals for the
$i$-th gauge group. The distance between the $i$-th and $(i+1)$-th
NS5 branes
yields the gauge coupling for the $U(N_i)$ gauge group while
coordinates of the D5 branes in the $x_7,x_8$ plane correspond to
the masses of fundamentals. The positions of D3 branes in the $x_7,x_8$
plane correspond to the coordinates on the Coulomb branch in the
quiver theory. The additional $\Omega$-deformation reduces the
theory with $N=4$ SUSY to the $N=2^{*}$ theory. At the energy scale
below the scale dictated by the lengths of the intervals the theory
on D3 branes is identified as $N=2^{*}$ 3d quiver gauge theory. In
what follows we assume that one coordinate is compact that is the 3d
theory lives on $R^2\times S^1$.

The mapping of the gauge theory data into the integrability
framework goes as follows. The Yang-Yang function is identified with
the twisted superpotential in the 3d gauge theory on the D3 branes and
its extrema yield solutions to the BE for the $XXZ$ spin chain or
equivalently the equation for the supersymmetric vacuum state in the
gauge theory \cite{ns1,ns2}. The D3 branes are identified with the
Bethe roots which are distributed according to the ranks of the
gauge groups at each of $n$ steps of nesting in  $\prod_{i}^{n}
U(N_i)$. Generically, the number of the Bethe roots at different
levels of nesting is different. The distances between the NS5 branes
define the twists at the different levels of nesting while the
positions of the D5 branes in the $x_7,x_8$ plane correspond to the
inhomogeneity parameters of the $XXZ$ spin chain.
To complete the dictionary,
recall that the anisotropy parameter
of the $XXZ$ chain is defined by the radius
of the compact dimension while the parameter of the $\Omega$-deformation
plays the role of the Planck constant for the $XXZ$ spin chain.
To get the $XXX$ chain from the $XXZ$ one,
one should just send the radius of the
compact coordinate to zero.

In terms of the brane configuration the dualities correspond to
particular brane motions. The bispectrality corresponds to the
interchange between the Coulomb and Higgs branches that is the mirror
symmetry \cite{gk}. To this aim, one should
adjust the parameters in such a way
that two D5 branes become at the same position in the 7,8,9 coordinates.
Then one should remove the segment of the D3 brane stretched between two D5
branes and bring it to infinity. The position of the D3 brane in the 7,8,9
coordinates
corresponds to the coordinate at the Higgs branch. Under the
bispectrality the roles of the D5 and NS5 branes get interchanged
and positions of the NS5 branes and D5 branes play the role of
inhomogeneity
parameters and twists respectively.

The interpretation of the QC duality is more
involved \cite{gk}. First, we have to perform the Hanany-Witten move
and translate all D5 branes to the left. Upon this move we get the
configuration involving the $\sum\limits_{i}K_i$
D5 branes yielding
the left boundary condition, $n$  NS5 branes defining the right
boundary condition and $Q$ D3 branes in between, where
 \beq\label{gc3719}
Q=\sum_{j=1}^n jK_j\,.
  \eeq

Since the distance between boundaries with the Dirichlet and Neumann
conditions is large, we get the $N=2^{*}$ $D=4$ gauge
theory with the $U(Q)$ gauge group on $R^2\times S^2\times I$. The
QC duality is now identified as the duality between the $N=2^{*}$
$D=3$ quiver gauge theory with particular content of fundamentals and the
$N=2^{*}$ $D=4$ theory with $U(Q)$ gauge group with nontrivial
boundary conditions. The information about the $D=3$ quiver is now
encoded in the boundary conditions of the $D=4$ theory via embedding
$SU(2)\rightarrow U(Q)$ at the left and right boundaries.

Now we are ready to explain the brane interpretation of the QC duality
in the degenerate XXX case we have elaborated.
At the spin chain side
the positions of  $n$ NS5 branes along the $x_3$
direction are identified with the
twist parameters $V_i$ and the number of the
NS5 branes fixes the rank of the
group. The positions of the D5 branes along $x_7,x_8$ are identified
with the inhomogeneity parameters $q_i$.
In the algebraic consideration given above we
considered the case when the total number of the $q_i$'s
(which is equal to $\sum_{j=1}^n jK_j$) coincides with the number of
particles $Q$ at the RS side. To get $Q=N$ we have
to put all $N$ D5 branes at the interval between two leftmost NS5
branes as it can be seen from (\ref{gc3719}). In this case $K_1= N$
and $K_j=0, j= 2,\dots n$.

Upon the Hanany-Witten move we get the $N=2^{*}$ $D=4$ $U(N)$ gauge
theory on the D3 branes. The object of interest in this theory is the
moduli space of the vacua which is known to be parameterized by the
$U(N)$ flat connections on the torus with one marked point with
particular holonomy determined by the $\Omega$-deformation parameter
\cite{nw,gw11}. This is exactly the description of the phase space of the
trigonometric RS model with $N$ particles \cite{gn}. One of the radii
of this auxiliary torus is the radius of the compact coordinate
which the 4d theory is defined on. Since we argued above that the
reduction from $XXZ$ to $XXX$ implies vanishing of this radius
in the 3d quiver theory, we have to take the same  limit in the 4d
theory. It can be immediately recognized as the transition from the
trigonometric to the rational RS model. Hence we arrive
exactly to the duality between the $XXX$ spin chain and the rational
RS model as it was discussed in \cite{gk}.

Now the boundary conditions fix two Lagrangian submanifolds in this
space. At the left Dirichlet boundary there are $N$ D5 branes which
provide the coordinates for the RS model with $N$ degrees of freedom
and correspond to $SU(N)$ holonomy around the cycle with the
vanishing radius. The second $SU(N)$ holonomy (around the cycle with
non-vanishing radius) corresponds to the Neumann  boundary conditions
imposed by the NS5 branes. Due to the nontrivial monodromy around
the marked point two holonomies can not be diagonalized simultaneously
and the second one can be identified as the Lax operator of the
rational RS model we have discussed above.
Hence we arrive at the picture
of intersection of two Lagrangian submanifolds. For the trigonometric case,
this picture has been discussed
for the first  time in
\cite{nrs}.

Now, the  algebraic consideration  of the previous section tells us
how the positions $V_i, i=1\dots n$, of the $n$ NS5 branes in the initial
quiver 3d  gauge theory (corresponding to the $GL(n)$ spin chain of length $N$)
provide the  multiplicity
of the Lax eigenvalues at the RS
side. Equation (\ref{gc502}) tells that $V_1$ has
multiplicity $N-N_1$, $V_2$ multiplicity $N_2-N_1$ etc. Since
$N_i$ is just the number of the D3 branes at the $i$-th segment, we could
claim that the structure of the clasterization of the Lax eigenvalues in
the RS model is dictated by the difference of the D3 branes at the
corresponding step of nesting.
Hence we obtain a very explicit
prescription how the quiver data in the 3d theory get mapped into the
choice of the particular Lagrangian submanifold in the moduli space
of vacua in the 4d gauge theory at the interval at small length of interval.

\section{Discussion}

In this paper we have described a clear-cut relationship between
the quantum $XXX$ spin chain and the rational
classical RS model. This QC duality and its
generalization to the trigonometric case has been discussed
in \cite{gk} in the brane framework but an explicit algebraic
analysis was missing. We put this on the firm ground and get
some important identifications. The spectrum
of the $XXX$ spin
chain Hamiltonians
coincides with possible values
of velocities of the RS particles under the conditions that
their coordinates equal the inhomogeneity parameters
of the spin chain and eigenvalues of the RS Lax matrix
coincide with the twist parameters with certain multiplicities
depending on the total ``spin projection''. The stationary
states of the quantum model appear to be in one-to-one
correspondence with intersection points of two Lagrangian
submanifolds in the phase space of the classical model.
The brane picture behind this
pattern has been presented.

This paper together with \cite{gk} has just
started the systematic investigation of the new type of
dualities in integrable models and their gauge theory meaning.
We believe that they may be
potentially very useful in many physical applications.
In particular, the possibility to find
spectra of quantum Hamiltonians in terms
of the $QC$-dual classical model seems to be
especially intriguing and promising.

We conclude by a list of
some interesting related topics deserving further investigation.

\begin{itemize}

\item The generalization of the algebraic analysis
to the trigonometric case is straightforward. We expect that
the QC duality extends also to integrable models with
elliptic $R$-matrices. However, such a generalization is going to be
non-trivial since in the elliptic case there are
no continuous twist parameters. This probably
means that they get quantized.

\item It would be extremely interesting to enrich the QC duality
by a recipe of finding, via the map to a classical system,
not only spectra of quantum Hamiltonians
but also the eigenstates themselves. We conjecture that such
specifically quantum information might be encoded in the
fine structure of the intersection of Lagrangian submanifolds.

\item A related
problem is to elucidate the meaning of the YY function
and Baxter's $Q$-operators in the context of
the CM-RS models. The YY function was conjectured \cite{nrs} to be
the generating function for Lagrangian submanifolds in the RS phase space.
However, the validity and consequences of this identification
deserve further study. As the results of \cite{zabrodin3} suggest,
the Baxter's $Q$-operators should be related to
Backlund transformations on the classical side. The details
are to be clarified.

\item It is important to extend our analysis to
the generalized duality suggested in \cite{gk}
when the number of inhomogeneity
parameters at the spin chain
side does not coincide with the number of particles at the RS side.
In the brane language this corresponds to the generic quiver.

\item The {\it quantum-classical} duality discussed in this paper
should be somehow extended to a {\it quantum-quantum} one,
when the classical CM or RS model gets quantized. The question is
what happens with the spin chains under this deformation.
Presumably, they turn into {\it non-stationary} models
described by equations of the Knizhnik-Zamolodchikov type.
This issue is also closely related to
evaluation of knot invariants \cite{bg2}.

\item So far only two types of the brane moves have been identified
as some dualities in the associated
integrable systems: the move corresponding to
Higgsing and the Hanany-Witten move.
    It would be interesting to obtain the dualities corresponding
    to more general brane moves.

\item Recently some new field theory generalizations of
higher rank Painlev\'e-Schlesinger equations and the
corresponding models of the Gaudin-Calogero type were suggested
\cite{AAMOZ2}. They should
respect the same kind of dualities and, therefore, it is tempting to
study possible continuous limits of spin chains
in order to find the dualities
between local and non-local models which may be of special interest.

\item As is mentioned in \cite{zz02}, the quantization of the spectral curves
of integrable chains leads to the relations (the Baxter equations)
which are very similar to their classical analogues. This might
allow one to interpret the QC duality in terms of a combination of
the spectral duality \cite{zotov02,zotov01} and the Symplectic Hecke
Correspondence \cite{LOZ,LOSZ} (cf. \cite{gn}).

\end{itemize}

\section{Appendix}

\setcounter{equation}{0}

Here we prove Propositions \ref{predl01} and \ref{predl02}.
\subsection*{Proposition \ref{predl01}}

In the appendix we employ the auxiliary notation ${\bf x}_N =
\{ x_i\}_N$, ${\bf y}_M =
\{ y_i\}_M$ for brevity. Another frequently used notation,
${\bf e}_N$, means the $N$-dimensional vector  $(1\,,\ldots\,,1)$,
so ${\bf x}_N-\hbar {\bf e}_N =\{x_i -\hbar \}_N$, etc.

Recall the statement: the pair of matrices
 \beq\label{gc5042}
 {\mathcal{L}}_{ij}({\bf x}_N,{\bf y}_M,g)=\frac{g\,\hbar}{x_i-x_j+
 \hbar}\prod\limits_{k\neq j}^N\frac{x_j-x_k+\hbar}{x_j-x_k}
\prod\limits_{\ga=1}^M\frac{x_j-y_\ga}{x_j-y_\ga+\hbar}\,,\ \ \
i\,,j=1\,,...\,,N
  \eeq
and
 \beq\label{gc5052}
{\widetilde {\mathcal{L}}}_{\al\be}({\bf y}_M,{\bf
x}_N,g)=\frac{g\,\hbar}{y_\al-y_\be+\hbar}\prod\limits_{\ga\neq
\be}^M\frac{y_\be-y_\ga-\hbar}{y_\be-y_\ga}
\prod\limits_{k=1}^N\frac{y_\be-x_k}{y_\be-x_k-\hbar}\,,\ \ \
\al\,,\be=1\,,...\,,M\,.
  \eeq
are related by the identity
 \beq\label{gc5062}
 \det_{N\times N}\big({\mathcal{L}}({\bf x}_N,{\bf y}_M,g)-\la\big)=(g-\la)^{N-M}
\det_{M\times M}\big({\widetilde {\mathcal{L}}}({\bf y}_M,{\bf
x}_N,g)-\la\big)\,.
  \eeq
To prove this, we need some technical lemmas.

\begin{lemma}\label{lem1} {\bf\cite{AASZ2}}
The matrices ${\mathcal{L}}$ and ${\widetilde {\mathcal{L}}}$ can be
represented in terms of diagonal matrices
 \beq\label{gc606}
 \begin{array}{c}\displaystyle{
 \mathcal{D}_{ij}=\delta_{ij}\prod\limits_{\ga=1}^M\frac{y_\ga-x_j}{y_\ga-x_j-\hbar}\,,}\qquad
i\,,j=1\,,...\,,N\,,
\\ \ \\
\displaystyle{
{\widetilde{\mathcal{D}}}_{\al\be}=
\delta_{\al\be}\prod\limits_{k=1}^N\frac{y_\be-x_k}{y_\be-x_k-\hbar}
\,,\qquad \al\,,\be=1\,,...\,,M\,,}
\end{array}
  \eeq
diagonal matrices $D_0$ and $D_\hbar$
 \beq\label{gc607}
 \begin{array}{c}
 (D_0)_{ij}({\bf u}_K)=\delta_{ij}\prod\limits_{k\neq
 i}^K(u_i-u_k)\,,\ \ \ (D_\hbar)_{ij}({\bf u}_K)=\delta_{ij}\prod\limits_{k\neq
 i}^K(u_i-u_k+\hbar)\,,\\ \ \\
i\,,j=1\,,...\,,K\,,
\end{array}
  \eeq
the Vandermonde matrix
$V_{ij}({\bf u}_K)=u_j^{i-1}$,
$i\,,j=1\,,...\,,K$,
and the triangular matrix
 \beq\label{gc609}
 \begin{array}{c}
(C_{\hbar,K})_{ij}= \left\{\begin{array}{l}\displaystyle{
\frac{(i-1)!\, \hbar^{i-j} }{(j-1)!(i-j)!}\,,}\ \ j\leq i\,,\\ \ \\
0\,,\ \ j>i\,,
\end{array}\right.\ \ \ i\,,j=1\,,...\,,K
\end{array}
  \eeq
in the following way:
 \beq\label{gc610}
 \begin{array}{c}
{\mathcal{L}}({\bf x}_N,{\bf y}_M,g)=g\,D^{-1}_\hbar({\bf x}_N)\,
V^T({\bf x}_N+\hbar\,{\bf e}_N)\,\left(V^T\right)^{-1}({\bf
x}_N)\,D_\hbar({\bf x}_N)\, \mathcal{D}=\\ \ \\
=g\,D^{-1}_\hbar({\bf x}_N)\, V^T({\bf
x}_N)\,C_{\hbar,N}^T\,\left(V^T\right)^{-1}({\bf x}_N)\,D_\hbar({\bf
x}_N)\, \mathcal{D}\,.
 \end{array}
  \eeq
 \beq\label{gc611}
 \begin{array}{c}
{\widetilde {\mathcal{L}}}({\bf y}_M,{\bf x}_N,g)=g\,D_0({\bf
y}_M)\, V^{-1}({\bf y}_M)\,V({\bf
y}_M-\hbar\,{\bf e}_M)\,D_0^{-1}({\bf y}_M)\, {\widetilde{\mathcal{D}}}=\\ \ \\
=g\,D_0({\bf y}_M)\, V^{-1}({\bf y}_M)\,C_{-\hbar,M}\,V({\bf
y}_M)\,D_0^{-1}({\bf y}_M)\, {\widetilde{\mathcal{D}}}\,.
 \end{array}
  \eeq
(here $(\ldots )^T$ means transposition of the matrix).
\end{lemma}

Notice that
$\det\mathcal{D}=\det\widetilde{\mathcal{D}}$.
Therefore, \underline{statement (\ref{gc5062}) can be rewritten
as}
 \beq\label{gc614}
 \det_{N\times N}\big({\mathcal{L}}_0({\bf x}_N,g)-\la\,\mathcal{D}^{-1}\big)=(g-\la)^{N-M}
\det_{M\times M}\big({\widetilde {\mathcal{L}}}_0({\bf
y}_M,g)-\la\,\widetilde{\mathcal{D}}^{-1}\big)\,,
  \eeq
where
 \beq\label{gc615}
 ({\mathcal{L}}_0)_{ij}({\bf
 x}_N,g)={\mathcal{L}}_{ij}\left.\right|_{M=0}=
 \frac{g\,\hbar}{x_i-x_j+\hbar}\prod\limits_{k\neq j}^N\frac{x_j-x_k+\hbar}{x_j-x_k}
\,,\ \ \ i\,,j=1\,,...\,,N
  \eeq
and
 \beq\label{gc616}
({\widetilde {\mathcal{L}}}_0)_{\al\be}({\bf y}_M,g)={\widetilde
{\mathcal{L}}}_{ij}\left.\right|_{N=0}=\frac{g\,\hbar}{y_\al-y_\be+\hbar}\prod\limits_{\ga\neq
\be}^M\frac{y_\be-y_\ga-\hbar}{y_\be-y_\ga}
\,,\ \ \ \al\,,\be=1\,,...\,,M\,.
  \eeq

 \begin{lemma}\label{lem2}
The l.h.s. of (\ref{gc5062}) (or that of (\ref{gc614})), i.e., the
function
 \beq\label{gc6161}
 |\,{\mathcal{L}}_N\,|(M)\stackrel{\hbox{\tiny{def}}}{=}
 \det_{N\times N}\big({\mathcal{L}}_0({\bf
x}_N,g)-\la\,\mathcal{D}^{-1}\big)
  \eeq
has no poles of the form $\frac{1}{x_a-x_b}$ or
$\frac{1}{x_a-x_b+\hbar}$ for all $ a\,,b=1\,,...\,,N$. All poles of
(\ref{gc6161}) come from the diagonal matrix
$\mathcal{D}^{-1}$.
 \end{lemma}

\vskip3mm\underline{\em{Proof}}:\vskip3mm

\noindent The idea is to represent
$|\,{\mathcal{L}}_N\,|(M)$ in the form of determinant of a matrix
whose elements have no poles of the form $\frac{1}{x_a-x_b}$ or
$\frac{1}{x_a-x_b+\hbar}$ for all $ a\,,b=1\,,...\,,N$. Using Lemma
\ref{lem1} we have
 \beq\label{gc6111}
 \begin{array}{c}
\det\big({\mathcal{L}}_0-\la\,\mathcal{D}^{-1}\big)=
\det\big({\mathcal{L}}_0^T-\la\,\mathcal{D}^{-1}\big)\\
\ \\ =\det\big(gD_\hbar V^{-1}C_{\hbar,N}V
D_\hbar^{-1}-\la\,\mathcal{D}^{-1}\big)=\det\big(gC_{\hbar,N}-\la
V\mathcal{D}^{-1}V^{-1}\big).
 \end{array}
  \eeq
The latter expression
does not contain any poles of the type $\frac{1}{x_a-x_b+\hbar}$.
However, it may contain poles  of the type $\frac{1}{x_a-x_b}$ since $\det
V=\prod\limits_{i>j}(x_i-x_j)$. Let us verify that all such poles
vanish if $\mathcal{D}$ is given by (\ref{gc606}). Indeed, the
inverse of the Vandermonde matrix is given by
 \beq\label{gc6112}
\left.
V^{-1}_{kj}=\frac{1}{(j-1)!}\,\p_\rho^{(j-1)}\prod\limits_{s\neq k}^N
\frac{\rho-x_s}{x_k-x_s}\right |_{\rho=0}\,.
  \eeq
Therefore, the matrix element $(V\mathcal{D}^{-1}V^{-1})_{ij}$ takes the
form
 \beq\label{gc6113}
\left. \big(V\mathcal{D}^{-1}V^{-1}\big)_{ij}= \sum\limits_{k=1}^N
V_{ik}\mathcal{D}^{-1}_{kk}V^{-1}_{kj}=\sum\limits_{k=1}^N
x_k^{i-1}\mathcal{D}^{-1}_{kk}\frac{1}{(j-1)!}\,\p_\rho^{(j-1)}\prod\limits_{s\neq
k}^N \frac{\rho-x_s}{x_k-x_s}\right|_{\rho=0}\,.
  \eeq
Consider the linear combination of columns $\displaystyle{
\sum\limits_{k=1}^N
x_k^{i-1}\mathcal{D}^{-1}_{kk}\prod\limits_{s\neq k}^N
\frac{\rho-x_s}{x_k-x_s}}=\sum\limits_{k=1}^N
\big(V\mathcal{D}^{-1}V^{-1}\big)_{ik}\rho^{k-1}$ which is a
generating function for them. Powers of the auxiliary variable
$\rho$ correspond to the values of $j-1=0\,,...\,,N-1$.  An
arbitrary pole $\frac{1}{x_a-x_b}$ appears in the sum for $k=a, b$.
The residue is given by
 \beq\label{gc6115}
\prod\limits_{s=1}^N(\rho-x_s)\left(\frac{x_a^{i-1}}{\rho-x_a}
\frac{\mathcal{D}^{-1}_{aa}}{\prod\limits_{l\neq
a\,, b}(x_a-x_l)}-\frac{x_b^{i-1}}{\rho-x_b}
\frac{\mathcal{D}^{-1}_{bb}}{\prod\limits_{l\neq
a\,, b}(x_b-x_l)}\right)\,.
  \eeq
This expression vanishes at $x_a=x_b$ if
$\mathcal{D}^{-1}_{aa}=\mathcal{D}^{-1}_{aa}(x_a)$. This is the case
of (\ref{gc606}). $\blacksquare$

 \begin{lemma}\label{lem3}
The r.h.s. of (\ref{gc5062}) (or that of (\ref{gc614})), i.e. the
function
 \beq\label{gc6163}
 |\,{\widetilde
{\mathcal{L}}}_M\,|(N)\stackrel{\hbox{\tiny{def}}}{=}\det_{M\times
M}\big({\widetilde {\mathcal{L}}}_0({\bf y}_M,g)-\la\,{\widetilde
{\mathcal{D}}}^{-1}\big)
  \eeq
has no poles of the form $\frac{1}{y_a-y_b}$ or
$\frac{1}{y_a-y_b+\hbar}$ for all $ a\,,b=1\,,...\,,M$. All poles of
(\ref{gc6163}) come from the diagonal matrix ${\widetilde
{\mathcal{D}}}$.
 \end{lemma}
The proof is similar to the previous Lemma \ref{lem2}.

\vskip5mm\underline{\em{Proof of Proposition \ref{predl01}}}:\vskip3mm

\noindent The proof is by induction in $M$. The nontrivial part of
$C_{\hbar,K}$ has a form of the left-justified Pascal's triangle (of
binomial coefficients) weighted by $\hbar^{i-j}$. Notice that
$\left(C_{\hbar,K}\right)_{jj}=1$ for all $j=1\,,...\,,K$.
Therefore,
 \beq\label{gc612}
 \det(g\,C_{\hbar,K}-\lambda)=(g-\lambda)^K\,.
  \eeq
Let us first check (\ref{gc5062}) for $M=0$ and arbitrary $N$ (or
$N=0$ and arbitrary $M$). Since
${\mathcal{D}}\left.\right|_{M=0}=\hbox{Id}_N$, it follows from
(\ref{gc610}) that
 \beq\label{gc618}
 \det({\mathcal{L}}_0-\lambda )=\det(g\,C_{\hbar,N}-\lambda)=(g-\lambda)^N\,.
  \eeq
Similarly,
$\det({\widetilde {\mathcal{L}}}_0-\lambda )=(g-\lambda)^M$.
Suppose (the induction assumption) that (\ref{gc614}) holds true for all
$N$ and some fixed $M-1$, i.e.,
 \beq\label{gc620}
|\,{\mathcal{L}}_N\,|(M-1)=(g-\la)^{N-M+1}|\,{\widetilde
{\mathcal{L}}}_{M-1}\,|(N)\,,\ \ \forall\, N\,.
  \eeq
In order to prove that (\ref{gc614}) holds for $M-1\to M$, we expand
both sides of (\ref{gc614}) as sums over poles in $y_M$ and compare the
results.

\noindent 1). Consider first the l.h.s. of (\ref{gc614}):
 \beq\label{gc621}
 \begin{array}{c}\displaystyle{
|\,{\mathcal{L}}_N\,|(M)=\det\left|\left|\,
\frac{g\,\hbar}{x_i-x_j+\hbar}\prod\limits_{k\neq
j}^N\frac{x_j-x_k+\hbar}{x_j-x_k}
-\lambda\delta_{ij}\prod\limits_{\ga=1}^M\frac{y_\ga-x_j-\hbar}{y_\ga-x_j}
\right|\right|,}
 \end{array}
  \eeq
  where $i\,,j=1\,,...\,,N$.
Notice that $|\,{\mathcal{L}}_N\,|(M)$ is a rational function of the
$y_M$ with simple poles at $x_1\,,...\,,x_N$. Therefore, it can be
represented in the form
 \beq\label{gc622}
\left.
|\,{\mathcal{L}}_N\,|(M)=|\,{\mathcal{L}}_N\,|(M)\right|_{y_M=\infty}+
\sum\limits_{l=1}^N\frac{1}{y_M-x_l}C_l\,.
  \eeq
The first term equals
$\,{\mathcal{L}}_N\,|(M)\left.\right|_{{y_M}=
\infty}=|\,{\mathcal{L}}_N\,|(M-1)$.
To find $C_l$, let us note that the pole $\frac{1}{y_M-x_l}$
appears only in the $ll$-th component of the second (diagonal) term
of matrix (\ref{gc621}), $-\la{\mathcal{D}}^{-1}_{ll}$. Hence,
 \beq\label{gc624}
 C_l=\Delta_{ll}\left.\right|_{y_M=x_l}\,\res\limits_{y_M=x_l}
 \left(-\la{\mathcal{D}}^{-1}_{ll}\right)\,,
  \eeq
where $\Delta_{ll}$ is the principal minor of the matrix
${\mathcal{L}}_N$ obtained by removing the $l$-th column and the $l$-th row.
It is easy to see that
 \beq\label{gc625}
\Delta_{ll}\left.\right|_{y_M=x_l}=|\,{\mathcal{L}}_{N-1}^l\,|(M-1)\prod\limits_{j\neq
l}^N\frac{x_j-x_l+\hbar}{x_j-x_l}\,,
  \eeq
where the index $l$ in ${\mathcal{L}}_{N-1}^l$ emphasizes that its
argument is ${\bf x}_{N-1}=x_1\,,...\,,x_{l-1},x_{l+1}\,,...\,,x_N$,
i.e. $\{{\bf x}_N\}\setminus x_l$. The residue in (\ref{gc624})
equals
$\hbar\la\prod\limits_{\ga=1}^{M-1}\frac{y_\ga-x_j-\hbar}{y_\ga-x_j}$.
Then expression (\ref{gc622}) takes the form
 $$
|\,{\mathcal{L}}_N\,|(M)=
 $$
 \beq\label{gc626}
=|\,{\mathcal{L}}_N\,|(M-1)+
\sum\limits_{l=1}^N\frac{\hbar\la}{y_M-x_l}
\prod\limits_{\ga=1}^{M-1}\frac{y_\ga-x_j-\hbar}{y_\ga-x_j}|
\,{\mathcal{L}}_{N-1}^l\,|(M-1)\prod\limits_{j\neq
l}^N\frac{x_j-x_l+\hbar}{x_j-x_l}\,.
  \eeq
By the induction assumption, the determinants
$|\,{\mathcal{L}}_N\,|(M-1)$ and $|\,{\mathcal{L}}_{N-1}^l\,|(M-1)$,
$l=1\,,...\,,N$ satisfy (\ref{gc614}), i.e.,
 \beq\label{gc627}
 \begin{array}{c}
|\,{\mathcal{L}}_N\,|(M-1)=(g-\la)^{N-M+1}|\,{\widetilde
{\mathcal{L}}}_{M-1}\,|(N)\,,\\ \ \\
|\,{\mathcal{L}}^l_{N-1}\,|(M-1)=(g-\la)^{N-M}|\,{\widetilde
{\mathcal{L}}}_{M-1}\,|(N-1)_l
 \end{array}
  \eeq
The lower index $l$ in the r.h.s. again emphasizes that the set of
its arguments is
 ${\bf x}_{N-1}=\{{\bf x}_N\}\setminus x_l$.

\vskip3mm

\noindent 2). The r.h.s. of (\ref{gc614}) is determined by
 \beq\label{gc630}
|\, {\widetilde {\mathcal{L} }}_{M}|\, (N)=
\det \left |\left |\,\frac{g\,\hbar}{y_\al-y_\be+\hbar}
\prod\limits_{\ga\neq \be}^M
\frac{y_\be-y_\ga-\hbar}{y_\be-y_\ga}
-\la\delta_{\al\be}\prod\limits_{k=1}^N\frac{y_\be-x_k-\hbar}{y_\be-x_k}
\right | \right |\,,
\eeq
where $\al\,,\be = 1\,,...\,,M \,$.
As it follows from Lemma \ref{lem3}, $|\,{\widetilde
{\mathcal{L}}}_{M}\,|(N)$ has no poles of the type $\frac{1}{y_a-y_b}$
or $\frac{1}{y_a-y_b+\hbar}$. Therefore, similarly to (\ref{gc622}),
we have the decomposition
 \beq\label{gc631}
|\,{\widetilde {\mathcal{L}}}_{M}\,|(N)=|\,{\widetilde
{\mathcal{L}}}_{M}\,|(N)\left.\right|_{y_M=\infty}+\sum\limits_{l=1}^N\frac{1}{y_M-x_l}{\tilde
C}_l\,.
  \eeq
At $y_M=\infty$ the matrix ${\widetilde {\mathcal{L}}}_{M}$ takes
the form $\mat{{\widetilde {\mathcal{L}}}_{M-1}}{0}{0}{g-\la}$.
Hence,
 \beq\label{gc632}
|\,{\widetilde
{\mathcal{L}}}_{M}\,|(N)\left.\right|_{y_M=\infty}=(g-\la)|\,{\widetilde
{\mathcal{L}}}_{M-1}\,|(N)\,.
  \eeq
To find ${\tilde C}_l$, notice that all poles of the type
$\frac{1}{y_M-x_l}$, $l=1\,,...\,,N,$ are contained only in the
element $\left({\widetilde {\mathcal{L}}}_{M}\right)_{MM}$.
Therefore, ${\tilde
C}_l=\Delta_{MM}\left.\right|_{y_M=x_l}\,
\res\limits_{y_M=x_l}\left({\widetilde
{\mathcal{L}}}_{M}\right)_{MM}$. It is easy to see that
$\displaystyle{\res\limits_{y_M=x_l}\left({\widetilde
{\mathcal{L}}}_{M}\right)_{MM}=\la\hbar\prod\limits_{k\neq l}^N\frac{x_l-x_k-\hbar}{x_l-x_k}}$
and
 \beq\label{gc635}
\Delta_{MM}\left.\right|_{y_M=x_l}=|\,{\widetilde
{\mathcal{L}}}_{M-1}\,|(N-1)_l
\prod\limits_{\be=1}^{M-1}\frac{y_\be-x_l-\hbar}{y_\be-x_l}\,,
  \eeq
where $|\,{\widetilde {\mathcal{L}}}_{M-1}\,|(N-1)_l$ is defined
in (\ref{gc627}).
Finally, for (\ref{gc631}) we have
 \beq\label{gc636}
 \begin{array}{c}
|\,{\widetilde {\mathcal{L}}}_{M}|\,(N)\\ \
\\ \displaystyle{=(g-\la)|\,{\widetilde
{\mathcal{L}}}_{M-1}\,|(N)+\sum\limits_{l=1}^N\frac{\la\hbar}{y_M\!-\!x_l}
\prod\limits_{k\neq
l}^N\frac{x_l\!-\!x_k\!-\!\hbar}{x_l\!-\!x_k}\,|\,{\widetilde
{\mathcal{L}}}_{M-1}\,|(N-1)_l
\prod\limits_{\be=1}^{M-1}\frac{y_\be\!-\!x_l\!-\!\hbar}{y_\be\!-\!x_l}\,.}
 \end{array}
  \eeq

\vskip3mm

\noindent 3). At last,  compare (\ref{gc626}) with
(\ref{gc627}) and (\ref{gc636}). In this way we arrive at the
equality
 \beq\label{gc637}
|\,{\mathcal{L}}_N\,|(M)=(g-\la)^{N-M}|\,{\widetilde
{\mathcal{L}}}_{M}\,|(N)\,,\ \ \forall\, N
  \eeq
that finishes the proof. $\blacksquare$

\subsection*{Proposition \ref{predl02}:
The ``non-relativistic'' limit $\hbar\rightarrow 0$.}

The analogue of Lemma \ref{lem1} is

\begin{lemma}\label{lem4} {\bf\cite{AASZ2}}
The matrices 
%
 \beq\label{gc0176}
\begin{array}{c}\displaystyle{
 {\mathcal{L}}_{ij}({\bf x}_N,{\bf y}_M,\omega)=\delta_{ij}\left(\om+\sum\limits^N_{k\neq
i}\frac{\hbar}{q_i-q_k}+\sum\limits_{\ga=1}^M
\frac{\hbar}{\mu_\ga-q_i}\right)+(1-\delta_{ij})\frac{\hbar}{q_i-q_j}}\,,\\
\ \\ i\,,j=1\,,...\,,N
\end{array}
  \eeq
and
 \beq\label{gc0326}
\begin{array}{c} \displaystyle{
{\widetilde {\mathcal{L}}}_{\al\be}({\bf y}_M,{\bf
x}_N,\omega)=\delta_{\al\be}\left(
\om-\sum\limits^M_{\ga\neq\al}\frac{\hbar}{\mu_\al-\mu_\ga}
-\sum\limits^N_{k=1}\frac{\hbar}{q_k-\mu_\al}\right)+\left(1-\delta_{\al\be}\right)
\frac{\hbar}{\mu_\al-\mu_\be}}\,,\\
\ \\ \al,\be=1\,,...\,,M\,.
\end{array}
  \eeq
can be represented in terms of the diagonal matrices
 \beq\label{gc6065}
 \begin{array}{c}\displaystyle{
 \mathcal{D}_{ij}=\delta_{ij}\sum\limits_{\ga=1}^M\frac{\hbar}{\mu_\ga-q_i}\,,\ \ \
i\,,j=1\,,...\,,N\,,}
\\ \ \\
\displaystyle{
{\widetilde{\mathcal{D}}}_{\al\be}=\delta_{\al\be}\sum\limits^N_{k=1}\frac{\hbar}{\mu_\al-q_k}
\,,\ \ \ \al\,,\be=1\,,...\,,M\,,}
\end{array}
  \eeq
the diagonal matrix
 \beq\label{gc6075}
 (D_0)_{ij}({\bf u}_K)=\delta_{ij}\prod\limits_{k\neq
 i}^K(u_i-u_k)\,,\ \ \
i\,,j=1\,,...\,,K\,,
  \eeq
the Vandermonde matrix
 \beq\label{gc6085}
 V_{ij}({\bf u}_K)=u_j^{i-1}\,,\ \ \
i\,,j=1\,,...\,,K
  \eeq
and
 \beq\label{gc6095}
 \begin{array}{c}
(C_{0,K})_{ij}= \left\{\begin{array}{l}
j\,,\ \ \ i=j+1\,,\ i=2\,,...\,,K,\\ \ \\
0\,,\ \ \  \hbox{otherwise}
\end{array}\right.
\end{array}
  \eeq
in the following way:
 \beq\label{gc6107}
 \begin{array}{c}
{\mathcal{L}}({\bf x}_N,{\bf y}_M,\om)=

\om+\hbar\, D^{-1}_0({\bf x}_N)\, \p_z V^T({\bf x}_N+z\,{\bf
e}_N)\,\left(V^T\right)^{-1}({\bf x}_N+z\,{\bf
e}_N)\,D_0({\bf x}_N)+\mathcal{D}\\ \ \\
=\om+\hbar\, D^{-1}_0({\bf x}_N)\, V^T({\bf
x}_N)\,C_{0,N}^T\,\left(V^T\right)^{-1}({\bf x}_N)\,D_0({\bf x}_N)+
\mathcal{D}\,.
 \end{array}
  \eeq
 \beq\label{gc6117}
 \begin{array}{c}
{\widetilde {\mathcal{L}}}_{\al\be}({\bf y}_M,{\bf
x}_N,\om)=\om+\hbar\, D_0({\bf y}_M)\, V^{-1}({\bf y}_M-z\,{\bf
e}_M)\p_zV({\bf
y}_M-z\,{\bf e}_M)\,D_0^{-1}({\bf y}_M)+{\widetilde{\mathcal{D}}}\\ \ \\
=\om-\hbar\, D({\bf y}_M)\, V^{-1}({\bf y}_M)\,C_{0,M}\,V({\bf
y}_M)\,D^{-1}({\bf y}_M)+{\widetilde{\mathcal{D}}}\,.
 \end{array}
  \eeq

\end{lemma}
Proposition \ref{predl02} can be proved either directly  or by taking
the limit $\hbar\rightarrow 0$ together with the substitutions
$g:=\exp(\hbar\omega)\,,\ \ \ \la:=\exp(\hbar\la)$.
After taking the limit one should also rescale the variables as
 \beq\label{gc651}
x_i\ \rightarrow\ x_i/\hbar\,,\ \ i=1\,,...\,,N\,,\\  y_\ga\
\rightarrow\ y_\ga/\hbar\,,\ \ \ga=1\,,...\,,M\,.
  \eeq
Then the statement of Proposition \ref{predl02} (\ref{gc5067})
with the matrices (\ref{gc017}), (\ref{gc032}) follows  from
Proposition \ref{predl01} (\ref{gc506}) for the matrices
(\ref{gc504}), (\ref{gc505}). The relation between (\ref{gc504}),
(\ref{gc505}) and (\ref{gc017}), (\ref{gc032}) is given by
(\ref{gc410}) with $\eta=\hbar$. The matrices $C_{\hbar,K}$ and
$C_{0,K}$ are related in the same way.

\begin{small}

\end{small}

\end{document}